\documentclass{mn2e}
\def\spose#1{\hbox to 0pt{#1\hss}}
\def\approxgt{\mathrel{\spose{\lower 3pt\hbox{$\sim$}}\raise 2.0pt\hbox{$>$}}}
\def\approxlt{\mathrel{\spose{\lower 3pt\hbox{$\sim$}}\raise 2.0pt\hbox{$<$}}}
\def\XMM{{\it XMM-Newton }}

\def\ie{{\it i.e. }}
\def\eg{{\it e.g. }}
\def\etal{{\it et al. }}
\usepackage{graphics,graphicx}

\title[AGN Feedback in M\,87]
{Feedback under the microscope: thermodynamic structure and
AGN driven shocks in M\,87}
\author[E. T. Million \etal]
{\parbox[]{6 in}{E. T. Million$^1$, N. Werner$^1$, A. Simionescu$^1$, 
S. W. Allen$^1$,\\
P. E. J. Nulsen$^2$, A. C. Fabian$^3$, H. B\"ohringer$^4$, 
and J. S. Sanders$^3$\\
\footnotesize{\it
$^1$Kavli Institute for Particle Astrophysics and Cosmology, Stanford University, 382 Via Pueblo Mall, Stanford, CA 94305-4060, USA; \\
SLAC National Accelerator Laboratory, 2575 Sand Hill Road, Menlo Park, CA 94025, USA\\
$^2$Harvard-Smithsonian Center for Astrophysics, 60 Garden St., Cambridge, MA,
02138, USA\\
$^3$Institute of Astronomy, Madingley Road, Cambridge CB3 0HA\\
$^4$Max-Planck-Institut f\"ur extraterrestrische Physik, Giessenbachstr, 85748 Garching, Germany\\
}
}}
\begin{document}
\renewcommand{\thefootnote}{\arabic{\footnote}}
\maketitle
\begin{abstract}
\noindent
We present the second in a series of papers discussing the thermodynamic 
properties of M\,87 and the central regions of the Virgo Cluster
in unprecedented detail.
Using a deep {\it Chandra} exposure (574 ks), we present high-resolution
thermodynamic
maps created from the spectra of $\sim$16,000 independent
regions, each with $\sim$1,000 net counts.
The excellent spatial resolution of the thermodynamic maps reveals
the dramatic and complex temperature, pressure, entropy and metallicity
structure of the system. 
The `X-ray arms', driven outward from M\,87 by the central AGN, are prominent
in the brightness, temperature, and entropy maps.
Excluding the `X-ray arms', the diffuse 
cluster gas at a given radius is strikingly isothermal.
This suggests either that the ambient cluster gas, beyond the arms,
remains relatively undisturbed by AGN uplift, or that
conduction in the intracluster medium (ICM) is efficient along azimuthal
directions, as expected under action
of the heat-flux driven buoyancy instability (HBI).
We confirm the presence of a 
thick ($\sim40$ arcsec or $\sim3$ kpc) 
ring of high pressure gas at a radius of $\sim180$ 
arcsec ($\sim14$ kpc)
from the central AGN. We verify that this feature is associated with a 
classical shock front, with an average Mach number $M=1.25$.
Another, younger shock-like feature 
is observed at a radius of $\sim40$ arcsec 
($\sim3$ kpc) surrounding the central AGN, with an
estimated Mach number $M\approxgt1.2$. 
As shown previously, if repeated shocks occur every $\sim10$ Myrs, as suggested
by these observations, then AGN driven weak shocks could produce enough
energy to offset radiative cooling of the ICM.
A high significance enhancement of Fe abundance is observed at 
radii $350-400$ arcsec ($27-31$ kpc).
This ridge is likely formed in the wake of the
rising bubbles filled with radio-emitting plasma that drag cool, metal-rich
gas out of the central galaxy.  We estimate that at least $\sim1.0\times10^6$
solar masses of Fe has been lifted and deposited at a radius of
$350-400$ arcsec; approximately the same mass of Fe is measured
in the X-ray bright arms, suggesting that a single generation of 
buoyant radio bubbles may be responsible for the observed Fe excess 
at $350-400$ arcsec.

\end{abstract}


\section{Introduction}
The Virgo Cluster is the second brightest, extragalactic,
extended X-ray source in the $0.1-2.4$ keV ROSAT band, and the
closest galaxy cluster (16.1 Mpc; Tonry \etal 2001).
The central galaxy of the Virgo Cluster, M\,87, hosts an active galactic
nucleus (AGN) that exhibits compelling 
evidence for complex interactions with the surrounding 
cluster gas (\eg B\"ohringer \etal 1995; Young \etal 2002; 
Forman \etal 2005, 2007).
M\,87 is, therefore, an {\it excellent}
object with which to study AGN driven processes (\ie AGN feedback)
that affect the centers of galaxies and
galaxy clusters. 

The relatively cool, dense gas at the centers of many galaxy clusters 
emits copiously at X-ray wavelengths. In the absence of
a significant heat source, this gas would be expected to
cool quickly and promote significant star formation,
at rates an order of magnitude larger than observed
(see Peterson \& Fabian 2006 for a review).   
The energy required to offset this cooling is 
widely believed to come predominantly from the central
AGN, though the exact process or processes by which this occurs remains unclear
(see \eg McNamara \& Nulsen 2007). 
AGN inflate bubbles of relativistic radio plasma, which 
displace the X-ray emitting gas, forming clear cavities.
This process is expected to provide a 
significant source of heating and turbulence
(Churazov \etal 2001; see also \eg Br\"uggen \& Kaiser
2002; Kaiser 2003; Br\"uggen 2003; De Young 2003; Ruszkowski \etal 2004a,b;
Heinz \& Churazov 2005).
Sound waves and internal waves generated by cavity inflation
also provide a mechanism to heat the 
intra-cluster medium (ICM).
These waves have been observed in both the Perseus and Centaurus clusters 
(\eg Fabian \etal 2003; Fabian \etal 2006; Sanders \& Fabian 2007; 
Sanders \& Fabian 2008).
AGN induced shock activity is a further promising avenue to heat the ICM.
Clear indications of 
AGN-induced shocks are seen in several nearby galaxies and clusters, the 
best examples being Hydra A, Virgo, and Perseus clusters
(\eg Nulsen \etal 2005a; Simionescu \etal 2009; Forman \etal 2005, 2007; 
Fabian \etal 2003, 2006). 
AGN induced shocks in clusters are typically weak;  
Hercules A hosts the strongest observed AGN driven shock in a cluster with
a Mach number $M\sim1.65$ (Nulsen \etal 2005b).
These outbursts can also
be very energetic;  
MS\,0735+7421 boasts the highest observed energy with $\sim6\times10^{61}$ ergs 
associated with its shock (Gitti \etal 2007).
In galaxies and galaxy groups, 
these shocks can be much stronger with Mach numbers as high as
$M\sim8$ as observed in Centaurus A (Croston \etal 2009; 
Kraft \etal 2003).
In M\,87, Forman \etal (2005) identified shock fronts
associated with an AGN outburst $\sim1-2\times10^7$ years ago.  They argued
that shocks may be the most significant mechanism of heating 
the cool ICM near the core by the AGN. Forman \etal (2007) also 
measure temperature and density jumps that are consistent with 
a weak shock with Mach number $M\sim1.2$.
However, important uncertainties remain. For example, significant temperature
jumps are not seen across all putative shock fronts (\eg Graham \etal 2008).
Additionally, the observed pressure features are sometimes 
several kpc thick (Fabian
\etal 2006), complicating our understanding of how these processes
work. 

In M\,87, two X-ray bright `arms', 
discovered by Feigelson \etal (1987) using
the {\it Einstein Observatory}, are the most striking features driven by
AGN activity.  These arms extend to the southwest and east 
of the galactic center, are composed of cool gas, 
and are spatially coincident with extended
radio emission (B\"ohringer \etal 1995; Belsole \etal 2001; Molendi 2002).
B\"ohringer \etal (1995; see also Churazov \etal 2001) 
propose that these structures are the result of 
gas uplift in the wakes of bubbles of radio plasma buoyantly rising in
the hot ICM. 
However, the details of this process and the role of magnetic fields
are largely not understood.
Subsequent {\it Chandra} and \XMM observations of M\,87 have revealed
many additional complexities in the system 
(\eg Belsole \etal 2001; Matsushita \etal 2002; Molendi
2002; B\"ohringer \etal 2002; Sakelliou \etal 2002; Young \etal 2002; 
Forman \etal 2005, 2007; Simionescu \etal 2007, 2008; Werner \etal 2006).

Here, 
we present the second
in a series of papers studying in unprecedented detail
the thermodynamic properties of M\,87.  
Using an ultra-deep (574 ks) observation, we utilize fully the excellent 
spatial resolution of the {\it Chandra
X-ray Observatory}.
This paper focuses on the overall thermodynamic structure 
and Fe distribution of the galaxy
and surrounding cluster, and on the properties of the strongest shock features.
Other papers will study the 
detailed physics of the X-ray bright
arms and inner post-shock regions (Werner \etal 2010; hereafter Paper I) 
and the history of 
chemical enrichment (Million \etal 2010, in prep; Paper III).

The structure of this paper is as follows.  Section 2 describes the data 
reduction, basic imaging analysis, and spatially resolved spectroscopy.
Section 3 presents detailed thermodynamic maps for the cluster.
Section 4 presents results on the radial properties
of the ambient ICM to the north and south of the AGN; this 
excludes the X-ray bright arms, which introduce strong multi-temperature
structure into the profiles. We place 
important new constraints on the isothermality of the ambient
cluster gas.
Section 5 discusses the results, focusing on the
shock features at radius $r\sim180$ arcsec,
the isothermality of the ambient cluster gas, and a
high significance metallicity bump at $r\sim5-6$ arcmin.
Section 6 summarizes our conclusions.

Throughout this paper, we assume that the cluster lies at
a distance of 16.1 Mpc, for which the linear scale is
0.078 kpc per arcsec.

\section{X-ray Observations and Analysis}

\begin{figure*}
\scalebox{0.92}{\includegraphics{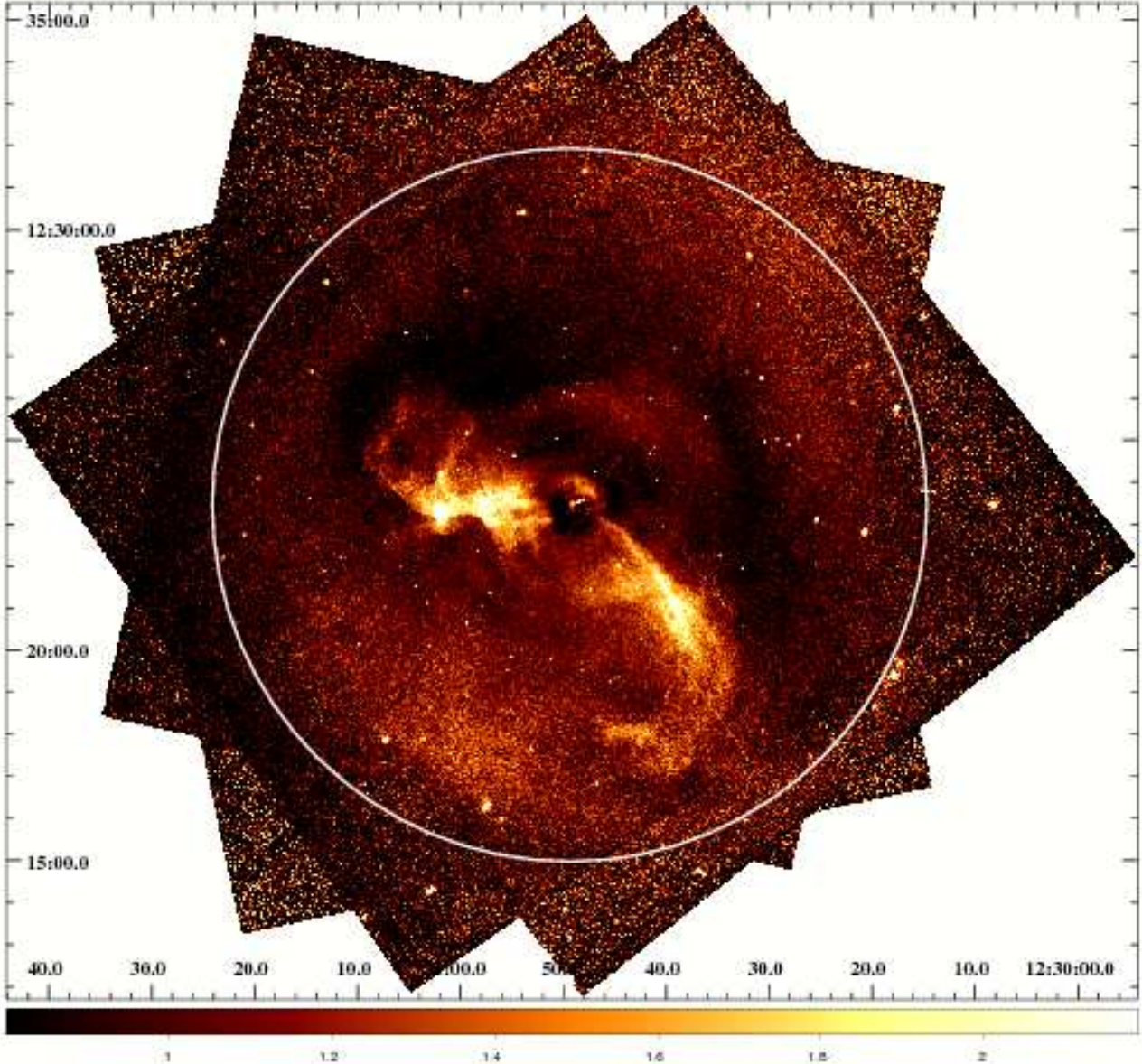}}
\caption{Exposure corrected, flat-fielded,  
surface brightness image of the central regions of the Virgo Cluster
divided by the best-fit, radially symmetric double-$\beta$ model
in the $0.6-2.0$ keV energy band and smoothed with a 2 arcsec Gaussian. 
Unity represents locations where the data is equal to the model.
This image uses only the ACIS-I data taken in 2005.
The white circle represents the outer extent of the thermodynamic maps
presented in Figs. \ref{fig:1kT}-\ref{fig:Z} (radius $8.5$ arcmin or $\sim40$ 
kpc).
}
\label{fig:sb}
\end{figure*}

\begin{figure*} 
\vspace{0.2cm}
\scalebox{0.92}{\includegraphics{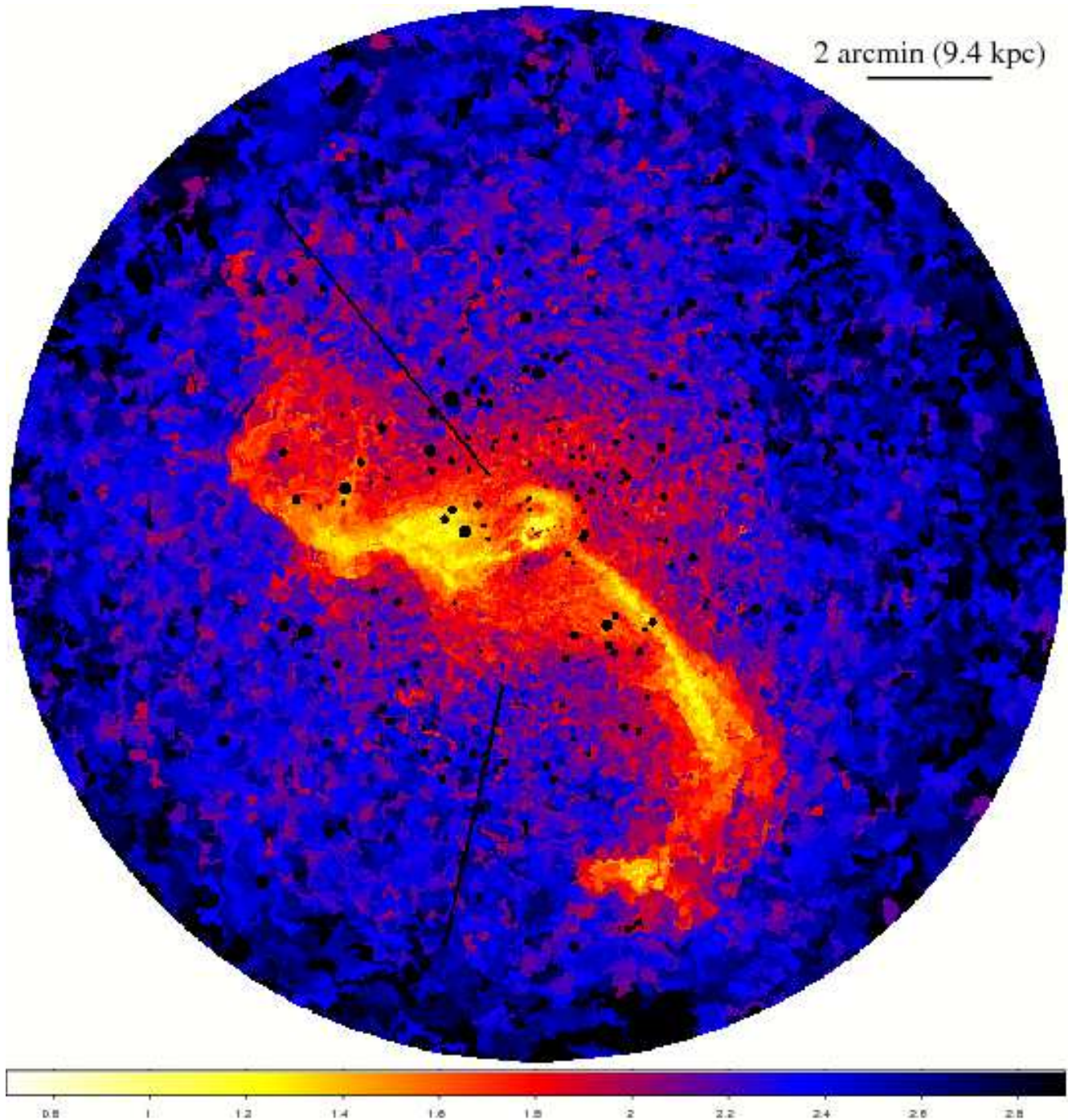}}
\caption{Temperature map of the central regions of the Virgo Cluster 
(in units of keV) for the 8.5 arcmin 
($\sim$40 kpc) radius
region indicated by the white circle in Fig. \ref{fig:sb}.
Regions have $\sim$1,000 net counts, leading to 1$\sigma$
fractional uncertainties in the plotted quantities of $\sim$10 per cent.
Point sources, and readout errors have been excluded
in this analysis and appear as black holes in the maps.
}
\label{fig:1kT}
\end{figure*}

\begin{figure*}
\vspace{0.2cm}
\scalebox{0.92}{\includegraphics{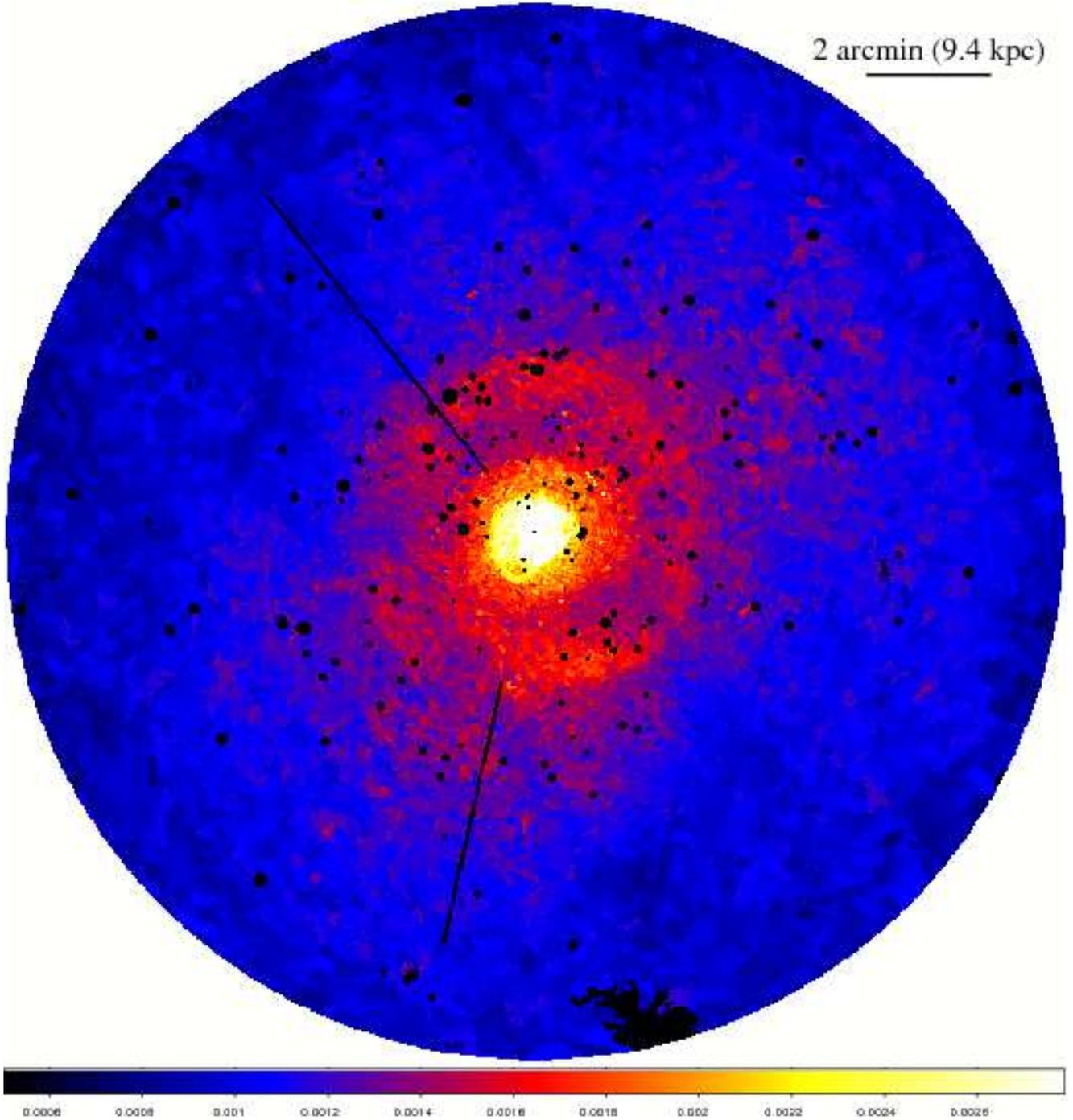}}
\caption{Pressure map in units of keV cm$^{-3}$ 
$\times \left(\frac{l}{2\,{\rm Mpc}}\right)^{-1/2}$.
The number density was calculated from the normalization of the spectra;
$n_e=\sqrt{\frac{C\,K}{A\,l}}$, where K is the {\small MEKAL}
normalization, A the projected area, and $l$ the line-of-sight depth.
$C=10^{14}\times 4\pi D_{\rm L}^2 /1.2$ and depends only on the 
distance to the source.
We assume that the 
line-of-sight depth is $l=2$ Mpc over the field of view.
Other details as for Fig. \ref{fig:1kT}.
}
\label{fig:P}
\end{figure*}

\begin{figure*}
\scalebox{0.92}{\includegraphics{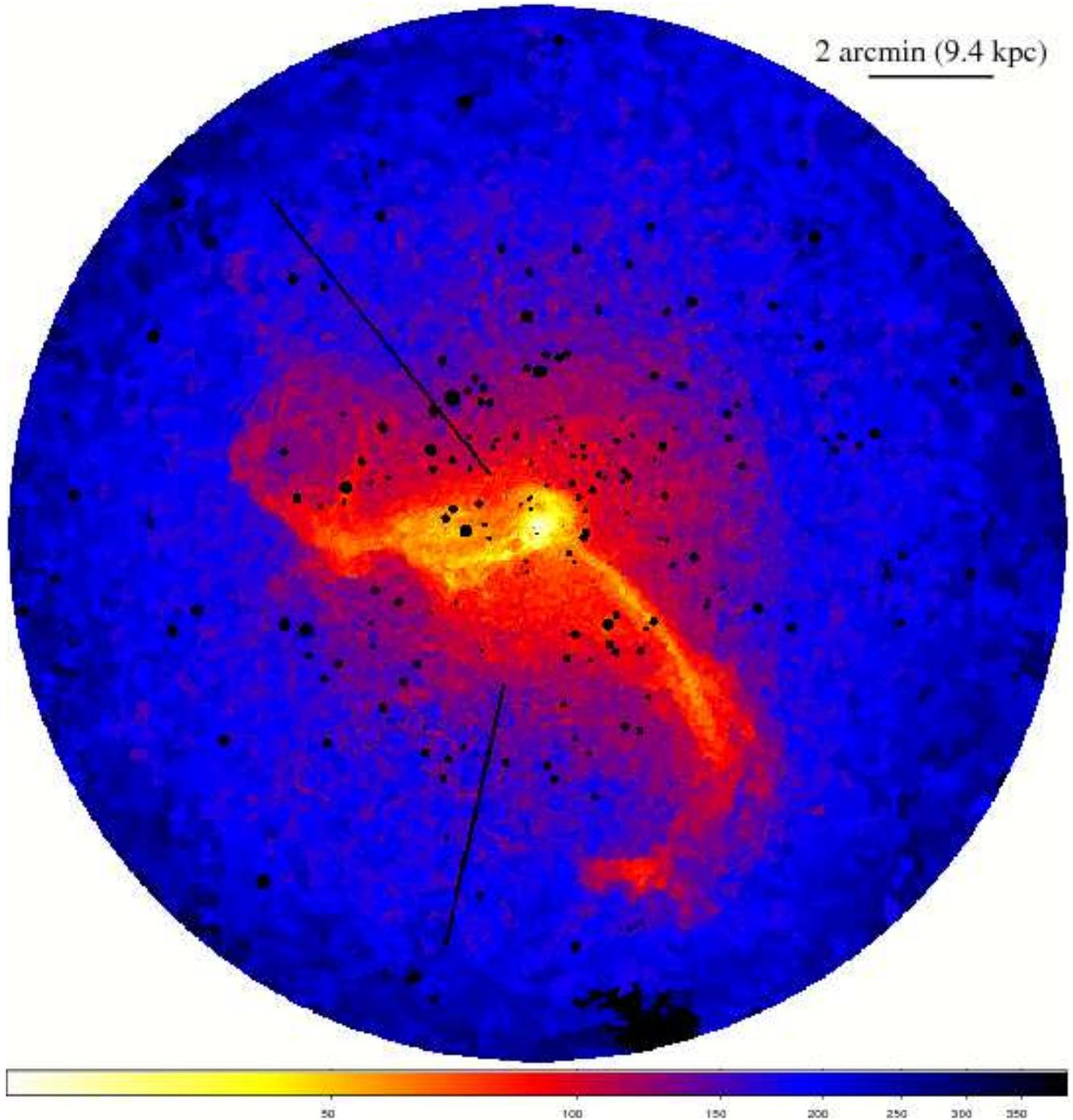}}
\caption{Entropy map of M\,87 in units of keV cm$^{2}$ $\times 
\left(\frac{l}{2 {\rm 
Mpc}}\right)^{1/3}$.
Other details as for Fig. \ref{fig:P}.
}
\label{fig:S}
\end{figure*}

\begin{figure*}
\scalebox{0.92}{\includegraphics{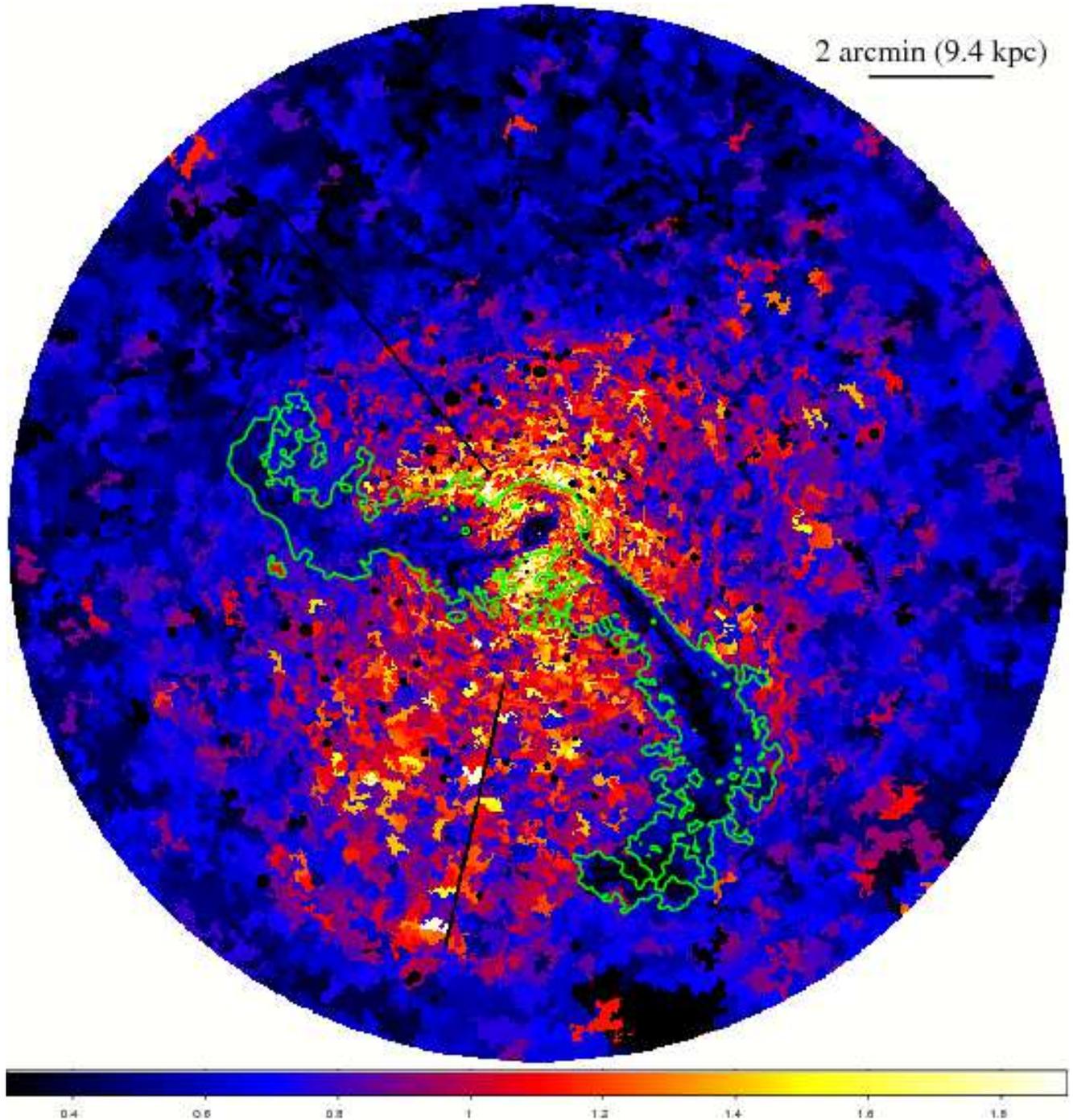}}
\caption{Metallicity map in solar units.  Abundances
are with respect to the proto-solar values of 
Lodders \etal (2003). Regions contain at least $2,500$ net counts
each. This results in statistical uncertainties of $\sim15$ per cent 
per region.
Regions inside the green contours contain significant 1 keV emission
(see Fig. \ref{fig:1kT};
Paper I) and are modelled improperly by a single temperature plasma.
As a result, the metallicity in these regions is significantly underestimated
(\eg Buote \etal 2000; Simionescu \etal 2008; Werner \etal 2008).
Other details as for Fig. \ref{fig:1kT}.
}
\label{fig:Z}
\end{figure*}

\begin{figure*}
\scalebox{0.43}{\includegraphics{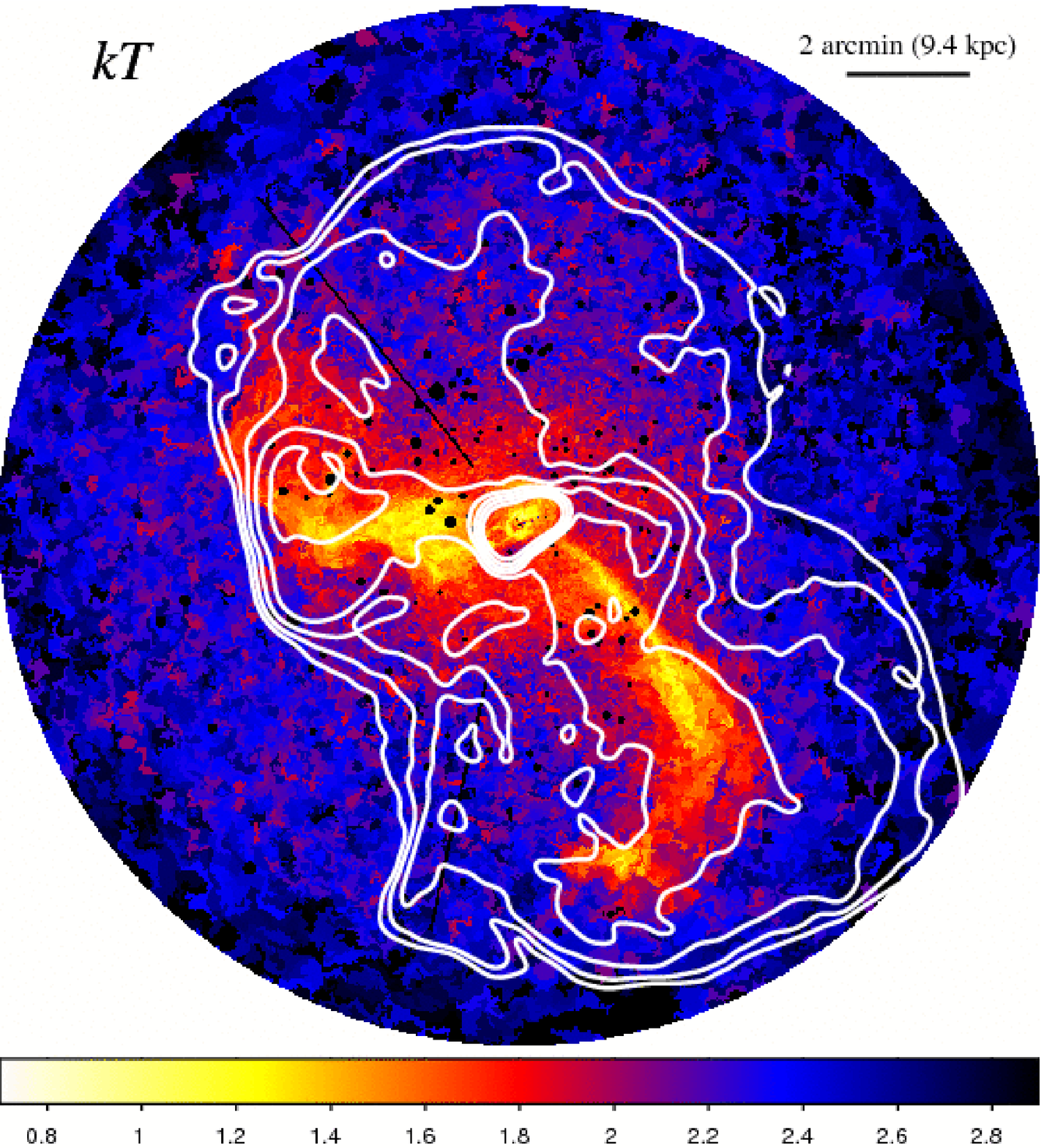}}
\hspace{0.9cm}
\scalebox{0.43}{\includegraphics{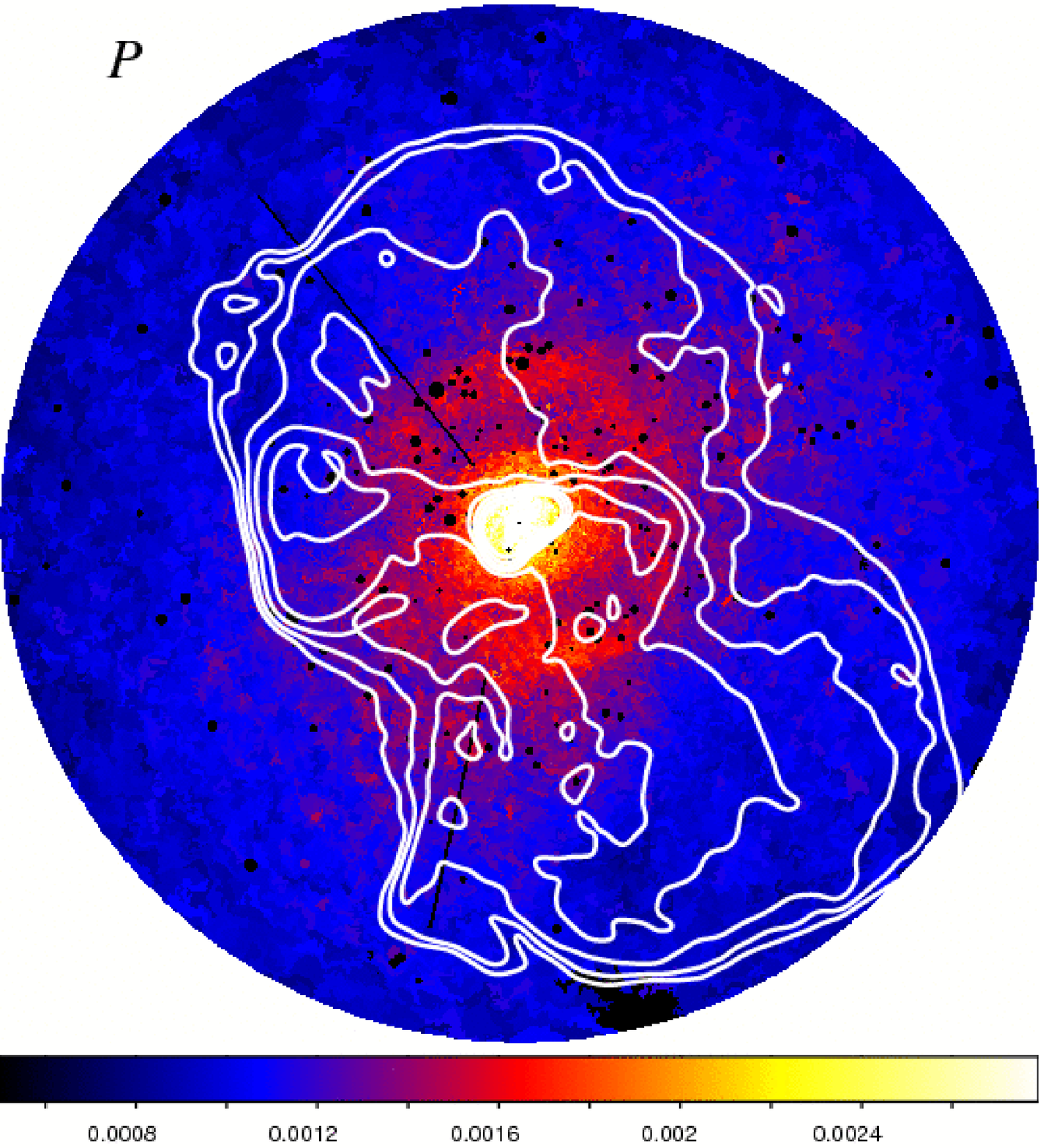}}\\
\vspace{0.3cm}
\scalebox{0.43}{\includegraphics{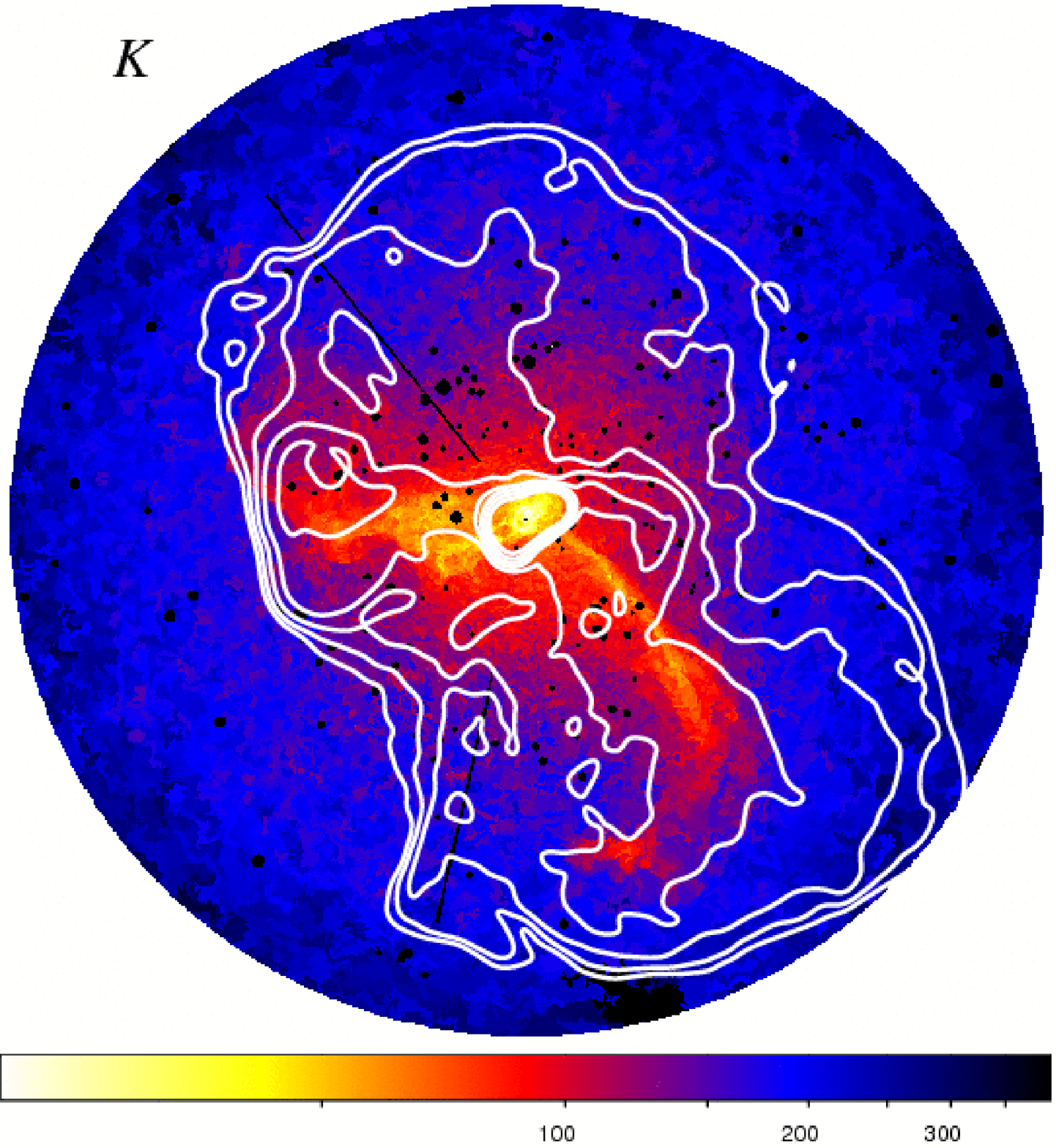}}
\hspace{0.9cm}
\scalebox{0.43}{\includegraphics{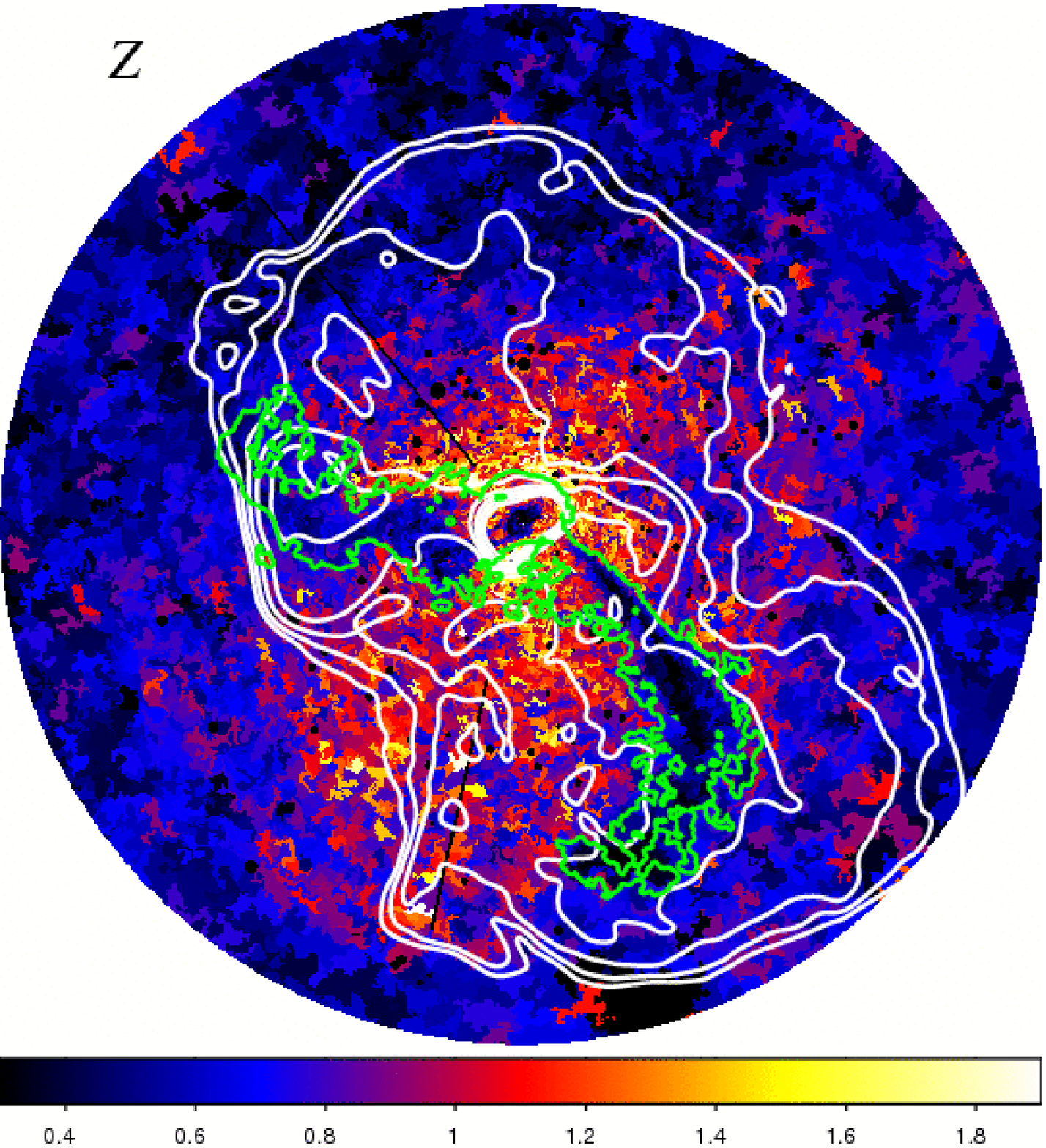}}
\caption{Temperature ($kT$), pressure ($P$), entropy ($K$), and metallicity 
($Z$) maps with the 
90 cm radio contours from Owen \etal (2000) overplotted.
Other details as for Fig. \ref{fig:1kT}--\ref{fig:Z}.
}
\label{fig:wradio}
\end{figure*}

\begin{table}
\begin{center}
\caption{Summary of the {\it Chandra} observations of M\,87 used in this study.
Columns list the 
observation ID, detector, observation mode, exposure after cleaning 
and observation date.}\label{table:obs}
\vskip 0 truein
\begin{tabular}{ c c c c c }
&&&&\\
Obs. ID & Obs. Date& Detector & Mode & Exp. (ks) \\
\hline \vspace{-0.2cm}\\
2707& Jul. 6  2002&ACIS-S&FAINT &82.9 \\
3717& Jul. 5  2002&ACIS-S&FAINT &11.1 \\
5826 & Mar. 3 2005&ACIS-I&VFAINT&126.8\\
5827 & May 5  2005&ACIS-I&VFAINT&156.2\\
5828& Nov. 17 2005&ACIS-I&VFAINT&33.0 \\
6186& Jan. 31 2005&ACIS-I&VFAINT&51.5 \\
7210& Nov. 16 2005&ACIS-I&VFAINT&30.7 \\
7211& Nov. 16 2005&ACIS-I&VFAINT&16.6 \\
7212& Nov. 14 2005&ACIS-I&VFAINT&65.2 \\
\hline
\end{tabular}
\end{center}
\end{table}

\subsection{Data Reduction}

Our work builds upon previous analyses of the same
datasets studied here (see Young \etal 2002; Forman \etal 2005, 2007).
The {\it Chandra} observations were carried out using the Advanced CCD
Imaging Spectrometer (ACIS) between July 2002 and November 2005.  
The standard level-1 event lists produced by the {\it Chandra} pipeline
processing were reprocessed using the $CIAO$ (version 3.5.0) software
package, including the appropriate gain maps and updated calibration 
products.
Bad pixels were removed and standard grade
selections applied. Where possible, the extra information available in
VFAINT mode was used to improve the rejection of cosmic ray
events. The data were cleaned to remove periods of anomalously high
background using the standard energy ranges 
recommended by the {\it Chandra} X-ray Center.  The net exposure times after
cleaning are summarized in Table~\ref{table:obs} and total 574 ks.
Separate photon-weighted response matrices
and effective area files were constructed for each region analyzed.

\subsection{Imaging analysis}

Background subtracted images on a $0.984\times0.984$ arcsec$^2$ pixel scale
were created for each pointing in many narrow 
energy bands (spanning $0.6-7.0$ keV).
These were flat-fielded with respect to the median energy for each image.
Background images were created from the blank-sky fields available from the
{\it Chandra} X-ray Center. The blank-sky fields were processed in an identical
manner to the science observations and were reprojected onto the same 
coordinates using the aspect solution files. The background images were
normalized by the ratio of the observed and blank-sky count rates in 
the $9.5-12$ keV band.
The images were summed to form a series of combined
background subtracted, exposure corrected 
images. 
Fig. \ref{fig:sb} 
shows the X-ray image in the $0.6-2.0$ keV band 
divided by the best-fit, double-$\beta$ model\footnote{
In detail, this is the sum of two $\beta$ models of the form 
$S_0\left[1+\left(\frac{r}{r_c}\right)^2\right]^{-\alpha}$,
where $S_0$ is the amplitude of the X-ray surface brightness, $r_c$ is the
core radius, and $\alpha$ is the power-law index.}
and smoothed with a 2 arcsec filtered Gaussian. 
This image spans the entire region covered by ACIS-I observations taken in 2005.
The white circle ($r\sim8.5$ arcmin) denotes the outer extent of the 
data used for the spectral analysis reported here.

The cool X-ray arms are readily apparent in the image.
An extensive imaging analysis of these data is presented by Forman \etal (2007).

\subsection{Spatially-resolved spectroscopy}

\subsubsection{Contour binning}

The individual regions for spectral fitting were determined using
the contour binning method of Sanders (2006), which groups
neighboring pixels of similar surface brightness until a desired
signal-to-noise threshold is met. For data of the quality discussed
here, regions outside the X-ray bright arms 
are small enough that the X-ray emission from
each can be approximated usefully by a single temperature plasma
model. Regions inside the X-ray bright arms show strong 
evidence for multi-phase
gas (see Paper I). In this paper, however, as a 
first approximation, we model all regions
as a single temperature plasma, recognizing that this will bias the results
for the X-ray bright arms.

We constrain each region to have
$\sim10^3$ net counts, giving $\sim$16,000 total regions for a total of
$\sim$16,000,000 net counts.  
This number of counts is sufficient to determine the temperature of the 
cluster gas to $\approxlt10$ per cent and the metallicity to $\sim$20 
per cent in each of the $\sim$16,000 independent spatial bins.  The density of
the cluster gas is always determined to higher precision.
We have imposed a lower limit on the region size of one arcsec$^2$, to 
mitigate the effects of {\it Chandra's} point spread function.
Regions vary in size from one to approximately 5,000 arcsec$^2$ in the outer
regions.
Because the statistical uncertainty of the best-fit metallicity is $\sim20$
per cent for regions with 1,000 counts, 
we have also performed a second analysis with regions containing
at least 2,500 net counts each. This latter 
analysis produces statistical uncertainties
of $3-5$ per cent on the temperature and $\sim15$ per cent on the metallicity.

\subsubsection{Annular binning}

To better study the properties of the ambient cluster gas, we have also
performed a radial analysis in which we exclude the X-ray bright arms.
This simplifies greatly the interpretation of the results.
For this analysis, we use annuli that vary in width from
10 to 30 arcsec, with $\sim$250,000 net counts in each for a total of 36 annuli.
Each annulus is fitted with a single temperature 
{\small MEKAL} or {\small APEC} spectral model 
(Section \ref{section:plasma}) with Galactic absorption.
We have analyzed the 
northern (position angles\footnote{All position angles are measured 
counter-clockwise from north.} -135 to 86 degrees) 
and southern (position angles 86 to 225 degrees) 
profiles both separately and together, always excluding the arms.

\subsubsection{Plasma models}
\label{section:plasma}

The spectra have been analyzed using XSPEC (version 12.5; Arnaud
1996). We use
the {\small MEKAL} and {\small APEC} plasma emission codes 
(Kaastra \& Mewe 1993; Smith \etal 2001), and the
photoelectric absorption models of Balucinska-Church \& McCammon
(1992). 
All spectral fits were carried out in the $0.6-7.0$
keV energy band. The extended C-statistic available in XSPEC, 
which allows for background subtraction, was used for all fitting.

The spectral model, applied to each spatial region, consists of
an optically-thin, thermal plasma model, 
at the redshift of the cluster.
We fix the Galactic absorption to $1.93\times10^{20}$ atom cm$^{-2}$ 
(determined from the $Leiden/Argentine/Bonn$
radio HI survey; Kalberla \etal 2005).
The temperature, metallicity, and the normalization 
are free parameters for every region (in detail, the normalizations
for each observation of each region are independently free).
In determining the metallicity from the 2-dimensional spectral maps,
all abundances are assumed to vary in 
a fixed ratio with respect to solar.
In our higher signal-to-noise analysis of annular spectra,
(see Section \ref{section:ambient}), we have allowed
the abundances of O, Ne, Mg, Si, S, Ar, Ca, Fe, and Ni to be 
separate, free parameters.
Results on the abundances of individual elements
will be presented in a subsequent paper (Paper III).
Abundances are given
with respect to the `proto-solar' values of Lodders (2003).  

\subsubsection{Background modelling}

Background spectra for the appropriate detector regions were extracted
from the blank-sky fields available from the {\it Chandra} X-ray
Center. 
The blank-sky fields were processed in an identical manner to the 
science observation
and were reprojected onto the same coordinates using the aspect
solution files.
Background regions were chosen to match the extraction regions
for the science observations.
The backgrounds were normalized by the ratio of the observed and
blank-sky count rates in the $9.5-12$ keV band (the statistical
uncertainty in the observed $9.5-12$ keV flux is less than 5 per cent
in all cases). 
Due to the very small effective area of the {\it Chandra} telescope at hard
X-ray energies, there is no significant cluster emission in the $9.5-12$
keV band.
Due to the high X-ray brightness of the target, uncertainties in the 
background modelling 
have little impact on the determination of thermodynamic quantities 
presented here.

\section{Thermodynamic Maps}

\begin{figure*}
\scalebox{0.43}{\includegraphics{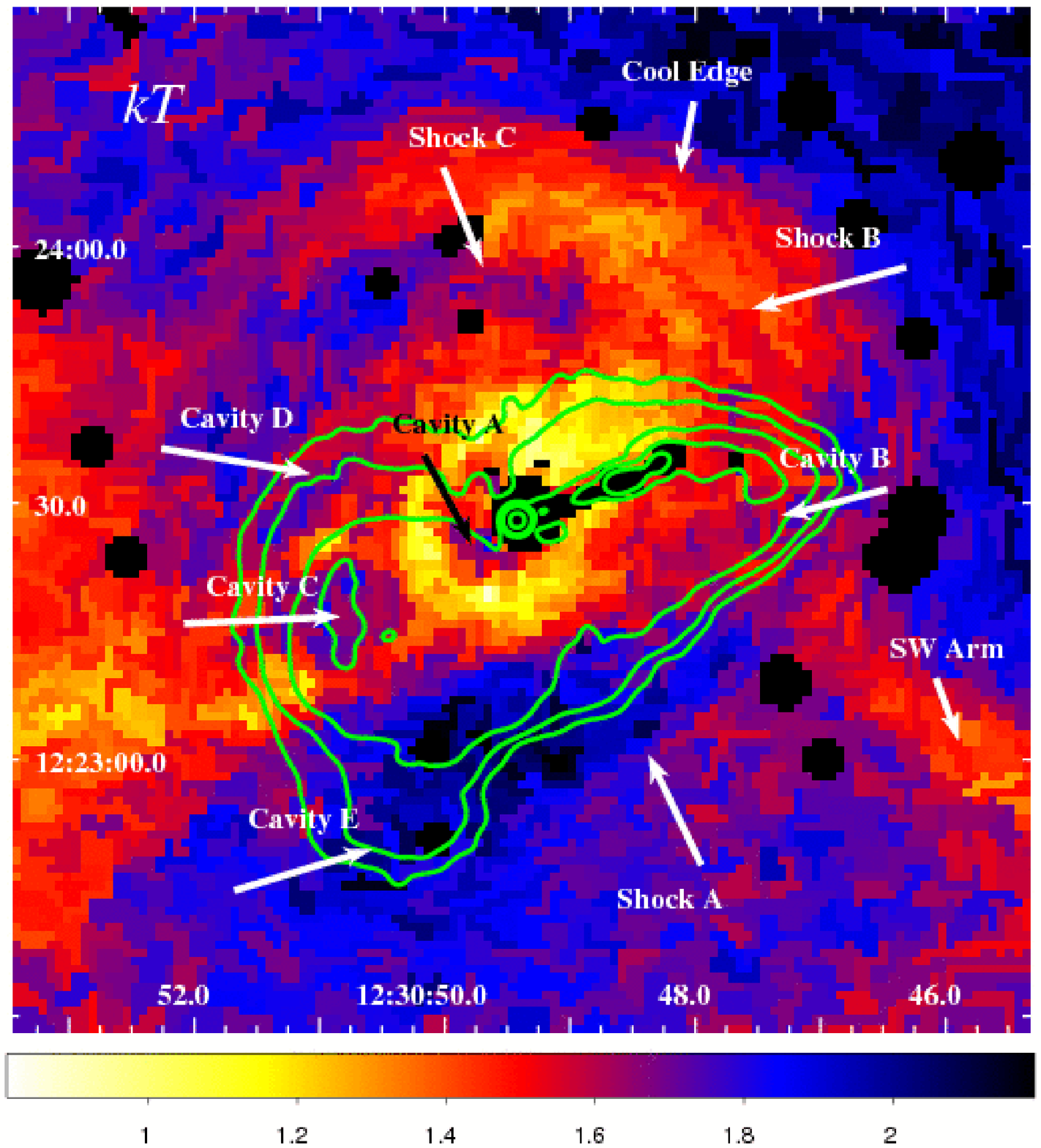}}
\hspace{0.9cm}
\scalebox{0.43}{\includegraphics{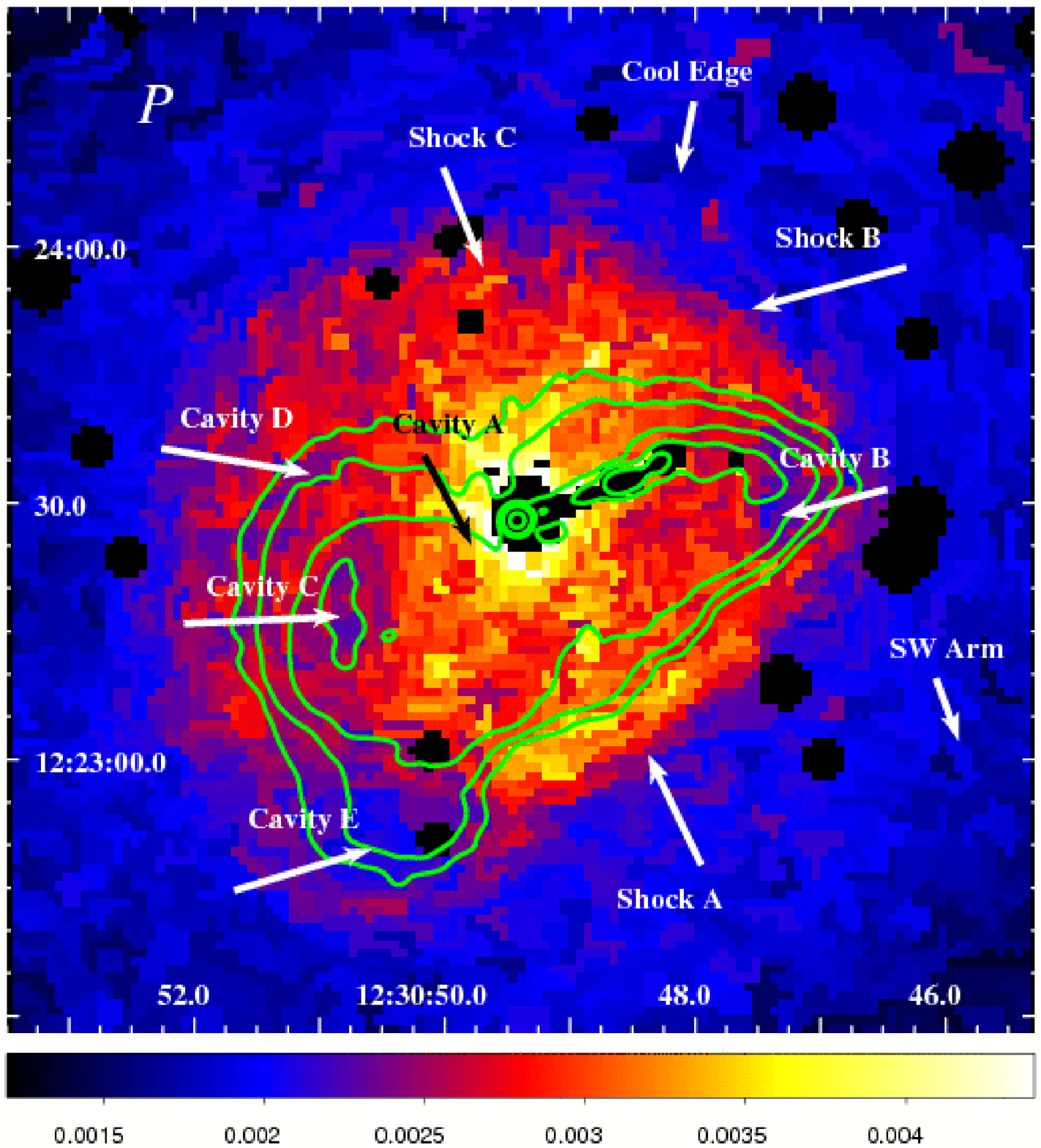}}
\caption{Thermodynamic maps (temperature on left; pressure on right)
of the central $2\times2$ arcmin$^2$ section of 
M\,87. Radio contours from Hines \etal (1989) are overlaid in green.
Several notable features are labelled with arrows.
}
\label{fig:zoom}
\end{figure*}

\subsection{Large-scale thermodynamic properties}
\label{section:thermo}

Figs. \ref{fig:1kT}--\ref{fig:Z} show the projected temperature,
pressure, entropy, and metallicity maps determined from our analysis.
Fig. \ref{fig:wradio} shows the same figures in mosaic form 
with the 90 cm (333 MHz) radio contours
from Owen \etal (2000) overlaid.
Point sources and read-out errors have been masked and ignored.
These results confirm spectrally, for the first time, 
many of the features identified by Forman \etal
(2005, 2007) from an imaging analysis of the same data-set.

The temperature map shown in Fig. \ref{fig:1kT}
dramatically reveals the complex temperature structure of the system.
The X-ray bright arms stand out as relatively cool structures
(Belsole \etal 2001; Young \etal 2002; Matsushita \etal 2002; Molendi 2002; 
Forman \etal 2005; Simionescu \etal 2007, 2008).  They 
are spatially correlated with enhanced 
90 cm radio emission (Fig. \ref{fig:wradio}a; contours are overlaid
in white; from Owen \etal 2000).  
However, within $\sim$3 arcmin (the radius of the shock front described in 
Section \ref{section:ring}), the southwestern arm does not coincide
precisely with the radio emission (see Fig. \ref{fig:wradio}a;
see also Forman \etal 2005, 2007).

Outside of the cool arms, the cluster 
appears remarkably isothermal at a given radius
(see also Section 5.1).  The average temperature,
excluding the arms, rises from $\sim2$ keV at a radius $r\sim0.9$ arcmin 
($\sim4$ kpc)
to $\sim2.7$ keV by the edge of the field ($r\sim8.5$ arcmin).
The X-ray bright, cold arms appear to be the most significant departures 
from spherical symmetry.  Surface brightness discontinuities 
at $r\sim3.75$ (northwest of the AGN) and $r\sim8$ 
arcmin (south of the AGN)
noted by Forman \etal (2005) do not have clear, corresponding features
in these maps.

The pressure map shown in 
Fig. \ref{fig:P} (Fig. \ref{fig:wradio}b 
with 90 cm radio contours from Owen \etal 2000) 
appears significantly more regular than the temperature map.
The exception is a ring of high pressure at a radius 
of $r\sim3$ arcmin that is approximately $\sim0.7$
arcmin thick. Forman \etal (2007) also see a feature at this radius
in their analysis and argue that it is consistent with 
an AGN driven shock propagating through the ICM with a Mach
number of $M\sim1.2$.

At the location of the cool arms, the observed pressure is somewhat lower.
This is, however, likely an artefact
due to the strong, projected multi-temperature structure in the direction of
the X-ray bright arms in these regions (Paper I).

The entropy map (Fig. \ref{fig:S}) reveals a similar morphological structure
to the temperature map.
The cool, X-ray bright arms have significantly
lower entropy than the ambient ICM at the same radius. 
This suggests that the gas within the
arms originates from the low entropy central regions of the cluster.  
The detailed thermodynamics of the arms are discussed in 
Paper I. 

The metallicity of the ambient 
cluster gas (Fig. \ref{fig:Z}) peaks at a value of
$\sim1.5$ solar near the core and drops to $\sim0.3$ near the edge of the field,
$r\sim8.5$ arcmin (40 kpc).  The southern region is also more metal rich
than the north and will be discussed further in Section 4.2.

At first glance, our maps suggest a reduced
metallicity within the X-ray bright arms with respect to the ambient gas
(the green contours of 1 keV emission; see Paper I). However, 
this apparent decrement is likely 
a bias due to the complex multi-temperature structure within the arms
(\eg Buote 2000; Simionescu \etal
2008; Werner \etal 2008 and references within). The metallicity
structure of the X-ray bright arms will be discussed further in Paper III.

A high metallicity structure is observed at a radius of
$r\sim4.5$ arcmin to the west of the central AGN.
This ridge connects with the high metallicity 
extension to the northwest of the core
at a radius of $r\sim3.3$ arcmin.
However, this ridge is not continuous, with some regions of low abundance
mixed within. 
As discussed in Section \ref{section:pancake}, this structure
is likely due to the uplift of cool, metal-rich gas in the wake of
buoyantly rising, radio-emitting plasma. High metallicity ridges
are also observed in the Perseus Cluster, Ophiuchus Cluster, and
Abell 2204 (see Sanders \etal 2005; Million \etal 2010; Sanders \etal 2009).

\subsection{Thermodynamic properties of the inner regions}

\begin{figure}
\scalebox{0.43}{\includegraphics{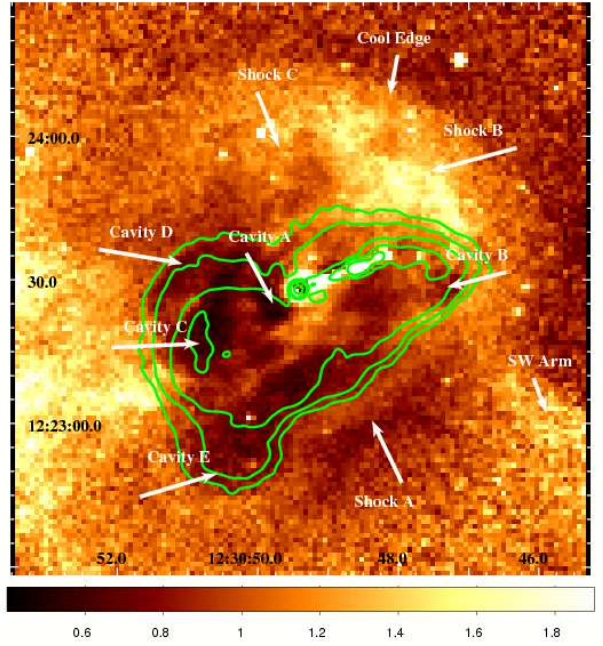}}
\caption{$0.6-2.0$ keV surface brightness image divided by an
azimuthally-symmetric double-$\beta$ model
of the central $2\times2$ arcmin$^2$ of Fig. \ref{fig:sb}.
Other details as for Fig. \ref{fig:zoom}. 
}
\label{fig:sbzoom}
\end{figure}

Temperature and pressure maps for the central $2\times2$ arcmin$^2$
($9.4\times9.4$ kpc$^2$)
region of M\,87 are shown on an enlarged scale in 
Fig. \ref{fig:zoom}. The 
6 cm (5 GHz) radio contours of Hines \etal (1989) have been overlaid.
Fig. \ref{fig:sbzoom} shows the $0.6-2.0$ keV residual X-ray image 
(extracted from Fig. \ref{fig:sb}) for the
same central $2\times2$ arcmin$^2$ region.
A number of prominent features in the images have been labelled with arrows.
Each labelled cavity is detected at more than the 
$10\sigma$ statistical significance in surface brightness with respect to
the smooth double-$\beta$ model.

The jets, which have a hard non-thermal spectral energy distribution (see
\eg Marshall \etal 2002; Wilson \& Yang 2002; Perlman \& Wilson 2005),
would normally
appear as bright, high pressure, hot structures in the maps, but have been
masked in these images.
At a radius of 7 arcsec (0.5 kpc) in the
counter-jet direction, we observe a low X-ray surface brightness, 
hot cavity (`A') (see also Young \etal 2002).
Interestingly, there is no obvious corresponding radio feature.
A region of brighter, cooler gas encircles this structure and the AGN.

Several other low surface brightness cavities
are visible in Fig. 8, in the 
jet and counter-jet directions (see also Forman \etal 2007).  These 
correspond to regions of apparent low pressure and high temperature.
Two of these cavities (labelled `B' and `C'; both are at a radius of 
$r\sim40$ arcsec from the nucleus in the jet and counter-jet directions)
are also filled with radio emitting plasma.  Rings of cooler gas surround
these cavities.  The eastern-most cavity (`D'), which also has relatively
low pressure, lies at the edge of the 
radio cocoon that surrounds the AGN.
Cavity E, labelled `bud' by Forman \etal (2005), 
is located $r\sim40$ arcsec (3 kpc) to the south of the AGN
and is associated with high temperature, low pressure gas
and is filled with radio plasma. 
Surrounding the cavity is a rim of high pressure material that is possibly 
shocked gas.

The southwestern edge of the bright radio emission corresponds to a high
pressure, high temperature, X-ray bright `ridge'.  
We identify this ridge as a shock front.
This feature (labelled `Shock A')
may extend to surround most of the bright, inner radio cocoon;
it is seen clearly to the northwest, though is less obvious to the east.
The connecting `bridge' 
\clearpage
\noindent between the central low temperature
region and the southwest arm, which is also cool, appears `interrupted' in
projection by this shocked gas.
To the northwest, for position angles of -63$-43$ degrees,
at a radius of $r\sim40$ arcsec from the AGN, a coherent arc of relatively
cool gas is observed. 
This cool gas is a multi-phase structure and is discussed 
in detail in Paper I.
The outer edge of this arc at $r\sim50$ arcsec 
(labelled `Cool Edge') is also probably physically linked to the central region 
of cooler gas, but is apparently broken up by the shock front (`Shock B'
and `Shock C').


\section{Ambient Gas Properties}
\label{section:ambient}

\subsection{Azimuthally averaged projected profiles}

Our thermodynamic mapping reveals, in exquisite detail, the complex
structure of the bright X-ray arms (see also Paper I).  
Just as importantly, {\it Chandra's} excellent
spatial 
\noindent resolution also allows us to mask these structures and
better determine the properties of the ambient gas surrounding M87.
In this manner, we can avoid the complicating effects of
temperature substructure, 
which will bias temperature and metallicity
results in cases where the wrong thermal model is assumed (\eg Buote 2000;
Simionescu \etal 2008; Werner \etal 2008).  

For this analysis, we use 36 annuli that vary in width from 10 to 30 arcsec,
each with $\sim250,000$ net counts.
Each region is fitted in the first case 
with a single temperature {\small MEKAL} spectral model, with Galactic
absorption. We also repeat the analysis using the {\small APEC} code.
The high signal-to-noise of our annular spectra 
allows us to constrain the abundances of 
O, Ne, Mg, Si, S, Ar, Ca, Fe, and Ni simultaneously, in addition to the
temperature and normalization of the cluster gas (19 free parameters;
17 free parameters at large radius).
Here we present results only for Fe. 
Results on elemental abundances
other than Fe will be presented in a subsequent paper (Paper III). 
Error bars are drawn at the $\Delta C=1$ (68 per cent) confidence level.

\begin{figure*}
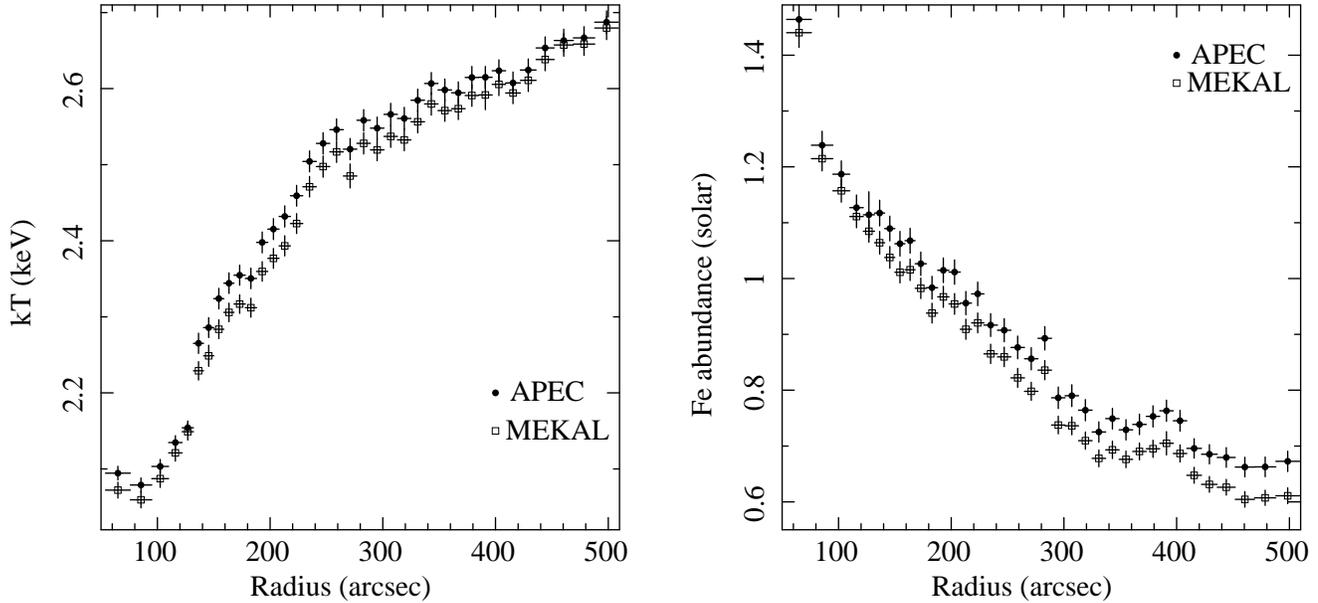

\hspace{0.0cm}
\scalebox{0.47}{\includegraphics[angle=270]{fig9a.ps}}
\hspace{0.4cm}
\scalebox{0.47}{\includegraphics[angle=270]{fig9b.ps}}
\caption{Azimuthally averaged, projected 
profiles of temperature (left; in keV) 
and Fe abundance (right) with respect to Lodders (2003) proto-solar 
abundances.
Results are shown for both the {\small MEKAL} (open squares) and 
{\small APEC} (filled circles) plasma emission codes (Kaastra \& Mewe 
1993, Smith \etal 2001, respectively). 
A radius of 100 arcsec corresponds to a physical scale of 
7.8 kpc. The X-ray bright arms are excluded from this analysis (see text).
The fits for these spectra include the abundances of O, Ne, Mg, Si, S, 
Ar, Ca, Fe, and Ni as separate, free parameters in addition to the 
temperature and normalization of the cluster gas.
}
\label{fig:kTFe}
\end{figure*}

Fig. \ref{fig:kTFe} 
shows the projected profiles of the temperature
and Fe abundance.
Results are shown for both the {\small APEC} (filled circles) and {\small MEKAL}
(open squares) plasma codes.
Our profiles
extend to the edge of the mosaic field of view, $r\sim8.5$ arcmin (40 kpc).
The inner radius for these profiles is 
defined by the `Cool Edge' identified at $r\sim50$ arcsec (4 kpc) in Fig. 
\ref{fig:zoom}a.

The temperature profile (Fig. \ref{fig:kTFe}a) steadily rises from
$\sim2.1$ keV to $\sim2.5$ keV for $50<r<240$ arcsec. Beyond $r\sim4$ arcmin
and 
out to $r\sim8.5$ arcmin, the temperature rises more slowly but again steadily,
reaching $kT\sim2.7$ keV by $r\sim8.5$ arcmin.
The statistical
errors on the measured temperatures range from 0.3 to 0.8 per cent.
The results for the two 
plasma codes exhibit slight differences, with the {\small 
MEKAL} temperatures being $\sim2$ per cent lower than those for {\small APEC}.  

Careful inspection of Fig. \ref{fig:kTFe}a reveals a 
jump in temperature at a radius of 
$r\sim130$ arcsec. This corresponds
to the inner edge of the shock front described in Section 
\ref{section:thermo}.
 
The Fe abundance profile (Fig. \ref{fig:kTFe}b) 
peaks at $Z_{\rm Fe}>1.4$ solar  
in the central regions. 
The profile then 
exhibits a steady decline to $Z_{\rm Fe}\sim0.7$ solar by $r\sim350$ arcsec.  
A significant enhancement, or `bump', in the 
Fe abundance is seen for radii $350<r<400$
arcsec ($27-31$ kpc).  This is approximately the radius
at which the bright X-ray arms terminate
(see also Section \ref{section:pancake}). 

We have investigated the impact of our choice of plasma code on the
measured Fe abundances, finding a systematic offset of $\sim2$ per cent 
in the results for the {\small MEKAL} and {\small APEC} codes at lower
temperatures.
This offset rises to $\sim10$ per cent for the hottest gas observed at the
edge of the field.

\begin{figure*}
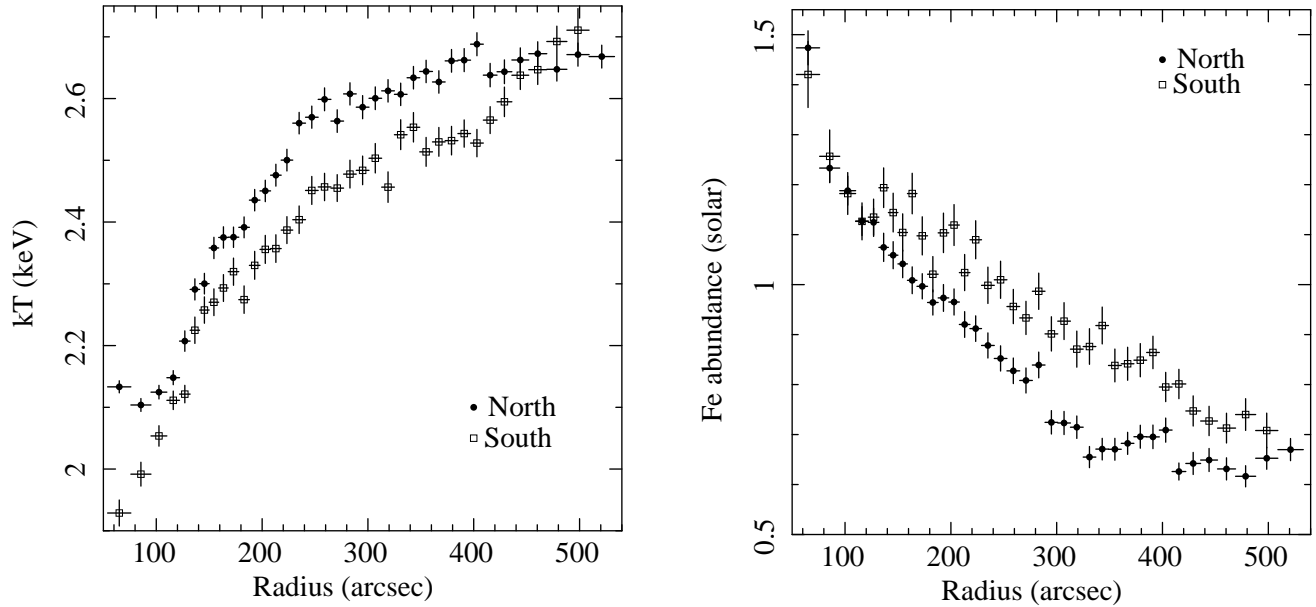

\hspace{-0.4cm}
\scalebox{0.47}{\includegraphics[angle=270]{fig10a.ps}}
\hspace{0.8cm}
\scalebox{0.47}{\includegraphics[angle=270]{fig10b.ps}}
\caption{Projected profiles of the temperature (left; in keV)
and Fe abundance (right; with respect to Lodders (2003) proto-solar 
abundances) for the northern and southern sectors, separately.
Here we use the {\small APEC} plasma emission code only.  Results for the 
northern sector (position angles -135 to 86 degrees) are plotted as 
filled circles.  Results for the
southern sector (position angles 86 to 225 degrees) are shown as open squares.
}
\label{fig:kTFeNS}
\end{figure*}

\subsection{Southern and northern profiles}

We have analyzed separately the 
properties of the ICM to the north and south
of the X-ray bright arms.\footnote{
The cool X-ray bright arms remain excluded in this analysis.}
Here, we have used the {\small APEC} plasma code only.
Fig. \ref{fig:kTFeNS} shows
the projected temperature and Fe abundance profiles to the north
(position angles -135 to 86 degrees; filled circles)
and south (position angles 86 to 225 degrees; open squares).  
Our results agree well with those of Simionescu \etal (2007) using \XMM data.

The northern gas temperature profile is approximately constant
at $kT\sim2.1$ keV for $50<r<100$ arcsec,
and then rises to $\sim2.7$ keV by 300 arcsec.  Beyond this, the temperature
rises only slightly out to $r\sim500$ arcsec.
In contrast, the southern profile shows a stronger temperature gradient.  The
central 100 arcsec dips below 2 keV
whilst at large radii $r\sim450$ arcsec the temperature rises to 
$kT\sim2.7$ keV.

The measured Fe abundance is consistently lower 
in the northern sector, presumably due to metal transport by sloshing 
(see Simionescu \etal 2010). 
For both the north and south, however, the Fe abundance peaks in the center and
drops steadily out to a radius $r\sim400$ arcsec.  To the north, we observe
a bump of enhanced Fe abundance at $330<r<400$ arcsec.
A less significant bump is also observed to the south.
This radius corresponds to the
outer radius of the X-ray bright arms.
A second large bump appears to begin at $r\sim500$ arcsec (39 kpc) to the north,
which corresponds to the outer radius of the overall radio halo
and suggests a similar origin. However, our profiles do not extend 
far enough to unambiguously confirm this feature.
The southern profile shows no convincing evidence of Fe abundance 
enhancements at $r\approxgt400$ arcsec.

\section{Discussion}
\subsection{Isothermality of the ambient cluster gas}
\label{section:iso}

The X-ray bright arms are well known to have multi-temperature structure
(\eg Belsole \etal 2001; Matsushita \etal 2002; Molendi 2002; 
Simionescu \etal 2008; Paper I).
Previous spectral analyses have suggested that the gas outside of the X-ray 
bright arms is approximately 
isothermal at a given radius (Matsushita \etal 2002).
Our data, however, allow us to study the spatial temperature
structure in more detail.
Such measurements have important implications for the physics
of the ambient ICM, including \eg the efficiency of thermal conduction
and the assumption of hydrostatic equilibrium.

To determine the presence of multi-temperature structure at a given radius,
we have made histograms of our different measurements of  
$kT$ in 0.3 arcmin wide annular regions
(Fig. \ref{fig:EMkT}).
The analysis was performed in eight annuli in the radial range of 
$90<r<235$ arcsec. For this analysis,  we use position angles of
-81--42 degrees. 
The width of the annuli was chosen to be large 
enough that $\sim75$ different
bins from the thermodynamic maps are present in each. For this analysis,
we use the temperature 
map created with at least $2,500$ net counts per region. 
The spectral models fitted to the individual spectra have the 
abundance ratios fixed to the best-fit values determined from
the analysis described in Section~\ref{section:ambient}.1 (see 
Section 5.1.1; see also Paper III). 

\begin{figure}
\scalebox{0.43}{\includegraphics{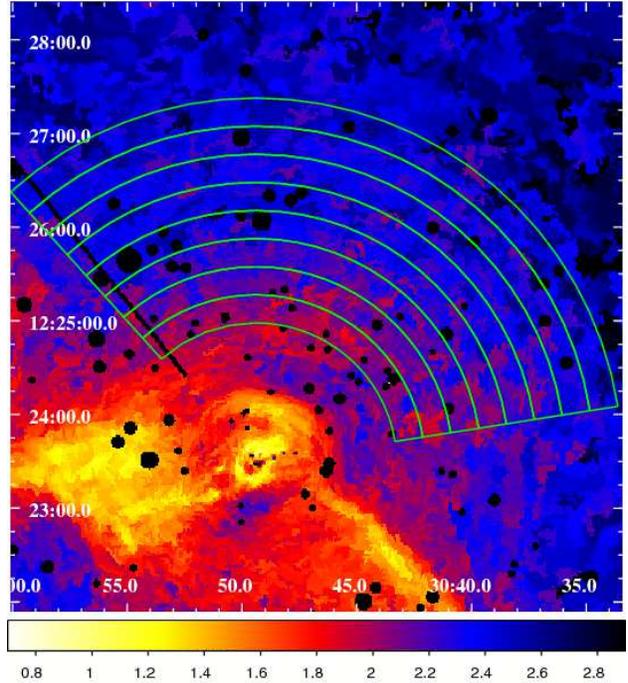}}
\caption{$6.5\times6.5$ arcmin$^2$ section of the
temperature map made with regions that contain $\sim2,500$ net counts.
The 8 partial annuli used in the analysis are overplotted in green.
The average error per bin $\sim5$ per cent in temperature. 
}
\label{fig:Z_iso}
\end{figure}

\begin{figure*}
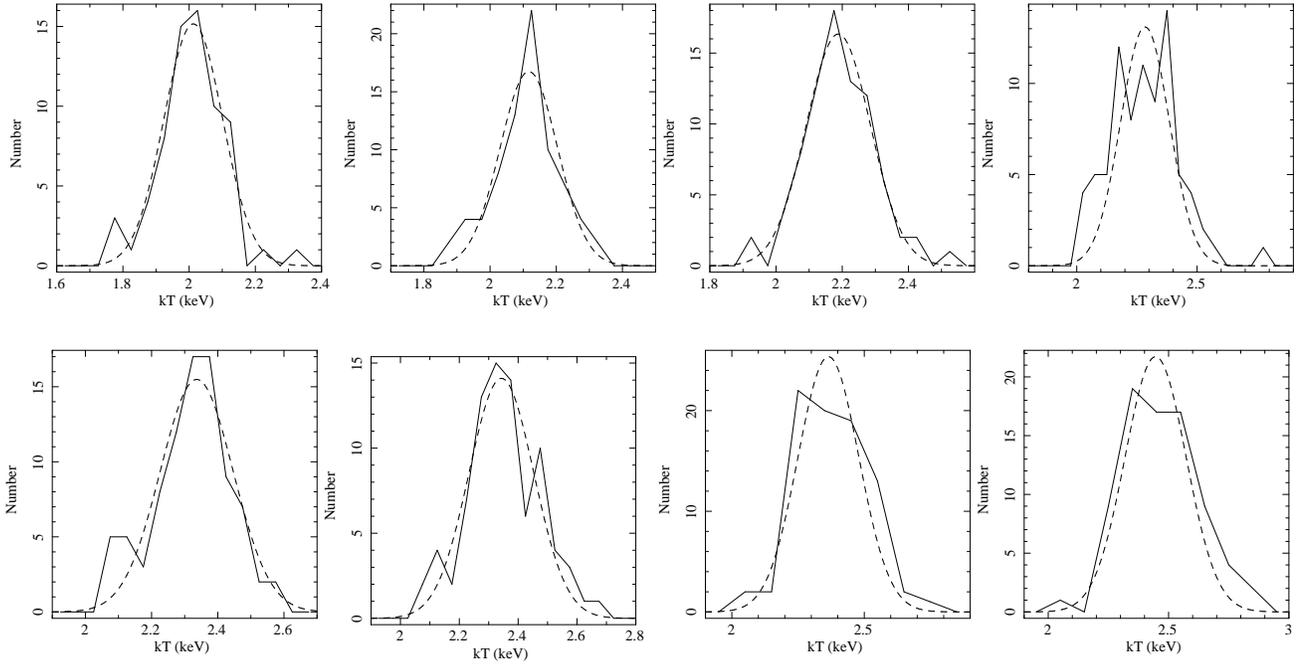

\hspace{-0.1cm}
\scalebox{0.24}{\includegraphics[angle=270]{fig12a.ps}}
\scalebox{0.24}{\includegraphics[angle=270]{fig12b.ps}}
\scalebox{0.24}{\includegraphics[angle=270]{fig12c.ps}}
\scalebox{0.24}{\includegraphics[angle=270]{fig12d.ps}}\\
\vspace{0.5cm}
\scalebox{0.24}{\includegraphics[angle=270]{fig12e.ps}}
\scalebox{0.24}{\includegraphics[angle=270]{fig12f.ps}}
\scalebox{0.24}{\includegraphics[angle=270]{fig12g.ps}}
\scalebox{0.24}{\includegraphics[angle=270]{fig12h.ps}}
\caption{Histograms of temperature measurements (in keV)
for all 8 regions. Each region contains $\sim75$ independent temperature
measurements. Gaussians with standard deviation equal
to that of the average measurement 
error per bin are overplotted (dashed lines). Only one region shows
significant deviation away from isothermality.}
\label{fig:EMkT}
\end{figure*}

We have also tested each individual region for spectral evidence of 
multi-temperature structure. 
A 4 temperature model (see Paper I) 
shows an absence of any significant thermal emission at temperatures 
between $0.5<kT<1.5$ keV at larger than the few per cent
level (see also Fig. \ref{fig:Z_iso}). 

\begin{figure}
\scalebox{0.47}{\includegraphics[angle=270]{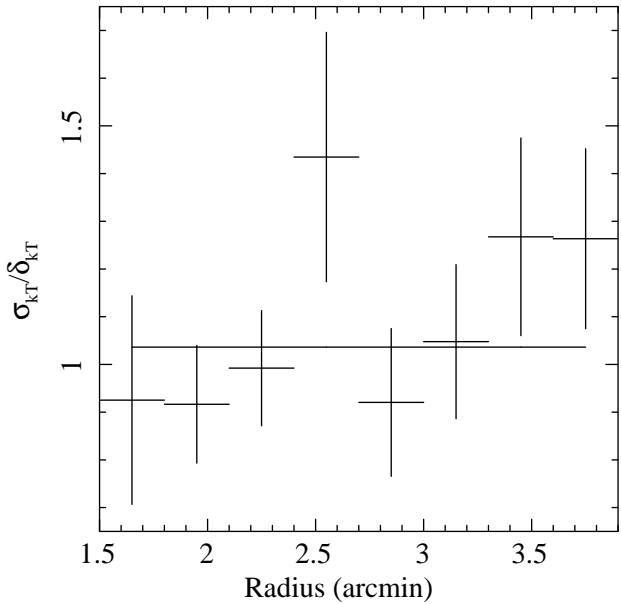}}
\caption{The ratio of the standard deviation of a Gaussian model 
fitted to the observed temperature distribution ($\sigma$) and the
average error per bin in the thermodynamic maps ($\delta$)
as discussed in Section \ref{section:iso}.
The ratio of errors is consistent with a constant value, which is 
overplotted on the data. 
}
\label{fig:errorratios}
\end{figure}

Having determined the distribution of 
temperature measurements, we compare this distribution with a Gaussian 
model. The width of the Gaussian is set to be equal to the average
statistical error per bin
(the typical standard deviations are approximately 3--5 per cent). 
Errors in each bin were determined from an MCMC analysis 
having chain lengths of at least 10$^4$ samples, after correcting
for burn-in. 
A good match between the temperature measurements and
the Gaussian model would be consistent with isothermality.
Fig. \ref{fig:errorratios} shows the ratio of the observed
width of the emission measure distribution (labelled $\sigma$) 
divided by the expectation, given the observed mean statistical error 
(labelled $\delta$). 

Only one out of 8 regions show evidence for a deviation from isothermality. 
This region is located at the 
temperature discontinuity behind the shock front. 
The temperature distribution in the other seven regions is consistent 
with being isothermal. 
Projection effects are not expected to have a significant influence on the
analysis.

The observed isothermality at a given radius implies either that the ambient
cluster gas beyond the arms is relatively undisturbed by AGN
uplift, or that there is efficient conduction within a spherical shell.
If conduction is responsible for the observed isothermality at a given radius,
then the likely orientation of the magnetic fields is perpendicular to the
radial direction, as expected under action of the heat-flux buoyancy
instability (Quataert 2008; see also \eg Bogdanovi\'c \etal 2009; Parrish 
\etal 2009).

Visual inspection of Fig. 5 also suggests that the metals in the ambient
ICM are significantly clumped. A detailed analysis of the metal distribution
and a discussion of the implications for turbulence and conduction in the ICM
will be presented in future work.

\subsubsection{Temperature bias}
\label{section:tbias}

\begin{figure*}
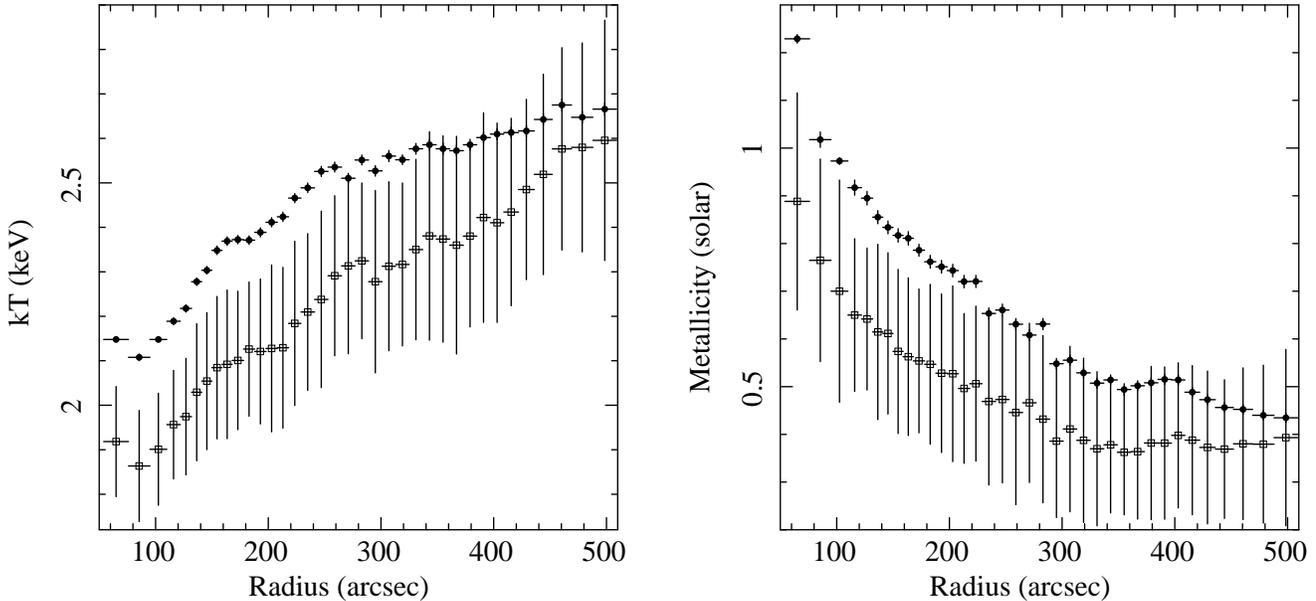

\scalebox{0.47}{\includegraphics[angle=270]{fig14a.ps}}
\hspace{0.4cm}
\scalebox{0.47}{\includegraphics[angle=270]{fig14b.ps}}
\caption{Temperature (left) and metallicity (right) 
measured from the mean of the 
emission measure distribution when fit with a Gaussian (open squares) and
from the integrated spectrum using 250,000 net counts and the same, simple
spectral model for the same annular regions in Section \ref{section:ambient}
(filled circles).
The spectral model is a single phase temperature model with elemental
abundances tied to Fe. Both the temperature and metallicity are systematically
offset by 10 and 30 per cent respectively when measured with the two techniques.
This is due to the incorrect modelling of the line emission around the Fe L
complex, where {\it Chandra's} effective area is largest.
}
\label{fig:bias}
\end{figure*}

Fig. \ref{fig:bias} shows the temperature (left) and metallicity (right) 
profiles measured from both the
peaks of the emission measure distributions
for the individual regions (open squares)
and from the integrated annular spectra presented in Section 
\ref{section:ambient} (filled circles).
The temperature and metallicity results from the two techniques
are systematically different 
by 10 and 30 per cent respectively.
The magnitude of this offset decreases as a function of radius.

From spectral simulations, 
we have determined that this bias is a combined modelling and
signal-to-noise issue.
Low temperature spectra (below $\sim3$~keV) contain a strong
line contribution at energies where {\it Chandra's}
effective area is the largest. However, in low signal-to-noise spectra, 
one cannot determine the abundances of individual elements separately.
These are instead
implicitly tied to the Fe abundance. The flux from line emission
is, therefore, modelled improperly which
biases the temperature and metallicity to low values. 
Increasing the signal-to-noise of the spectra reduces these biases, while
simultaneously causing the residuals around the Fe L complex to significantly
increase.  
The apparent position dependence is, therefore, explained by the temperature
dependence of the intensity of these line components.
The bias in low signal-to-noise spectra can be lessened
by fitting a model with the abundance ratios of the individual
elements adjusted to their best-fit values, as determined from a fit to 
the higher signal-to-noise
spectra.

We emphasize that this bias does not affect the primary 
conclusions of this paper.
Where appropriate, we have used the abundance ratios measured from the high
signal-to-noise 
radial analysis (see Paper III) to reduce the impact of this effect.

\subsection{Analysis of the shock fronts}
\subsubsection{The pressure ring at $r\sim180$ arcsec (14 kpc)}
\label{section:ring}

\begin{table}
\begin{center}
\caption{Summary of the shock models for each of the 4 different
surface brightness profiles. Columns list the position angles of
included data, shock radius, and Mach number of the observed shock
within each position angle range. Conservative uncertainties on the Mach
number, which bracket the observed scatter in the surface brightness
profiles, are estimated to be $M\pm0.04$.}
\begin{tabular}{ c c c }
&&\\
Position Angle (degrees) & $r$ (arcsec) & $M$\\
\hline \vspace{-0.2cm}\\
20--50 & 182 & $1.16$\\
190--220 & 168 & $1.29$\\
320--350 & 188 & $1.21$\\
350--20 & 180 & $1.34$\\
\end{tabular}
\end{center}
\label{table:shock}
\end{table}

The large-scale pressure map (Fig. \ref{fig:P}) reveals
the presence of a thick ($\sim40$
arcsec) high pressure ring at $\sim$180 arcsec.  This feature has previously
been identified as a shock with Mach number $M\sim1.2$ 
and age $\sim14$ Myrs (Forman \etal 2007).

\begin{figure*}
\scalebox{0.41}{\includegraphics{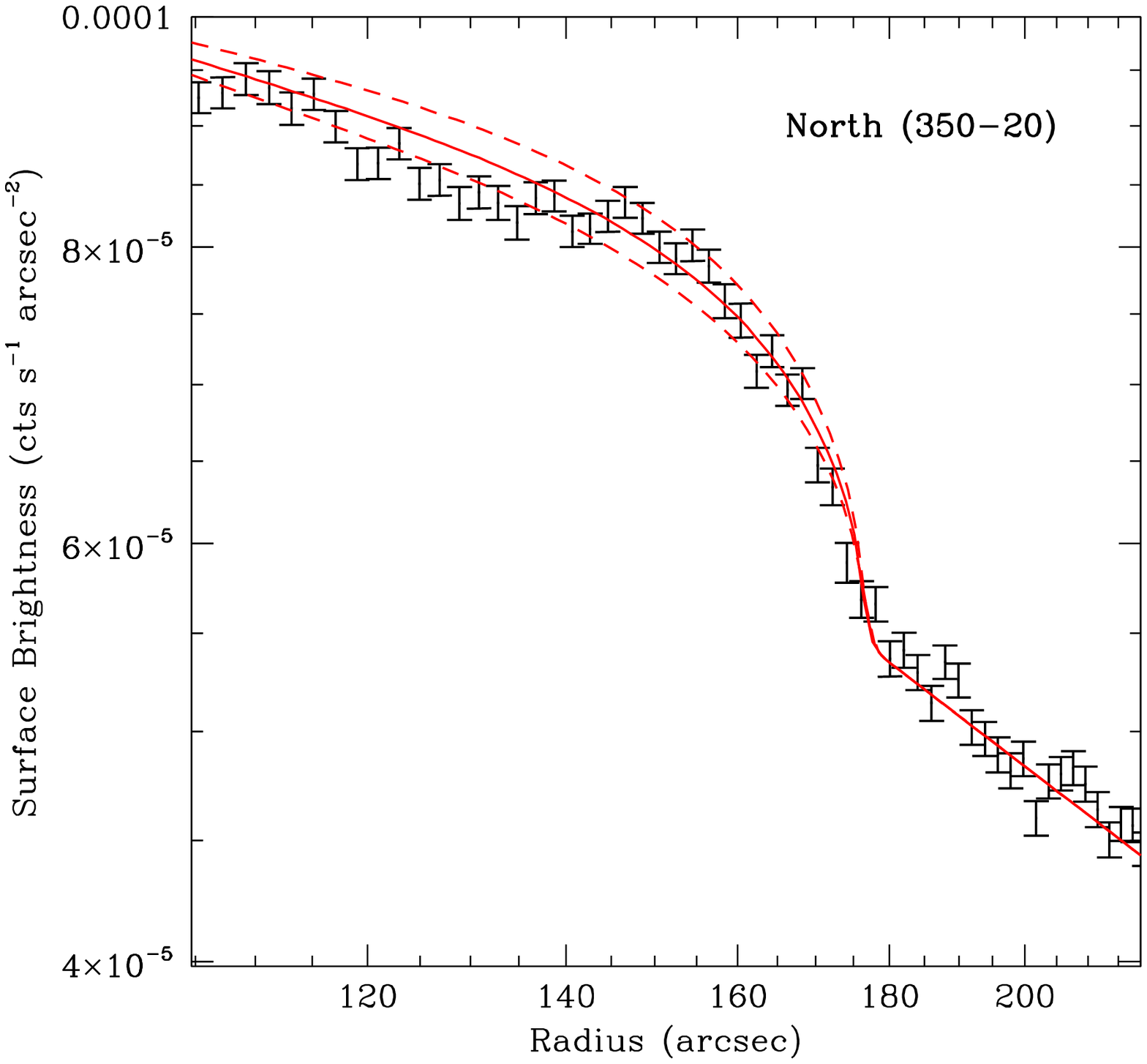}}
\hspace{0.5cm}
\scalebox{0.41}{\includegraphics{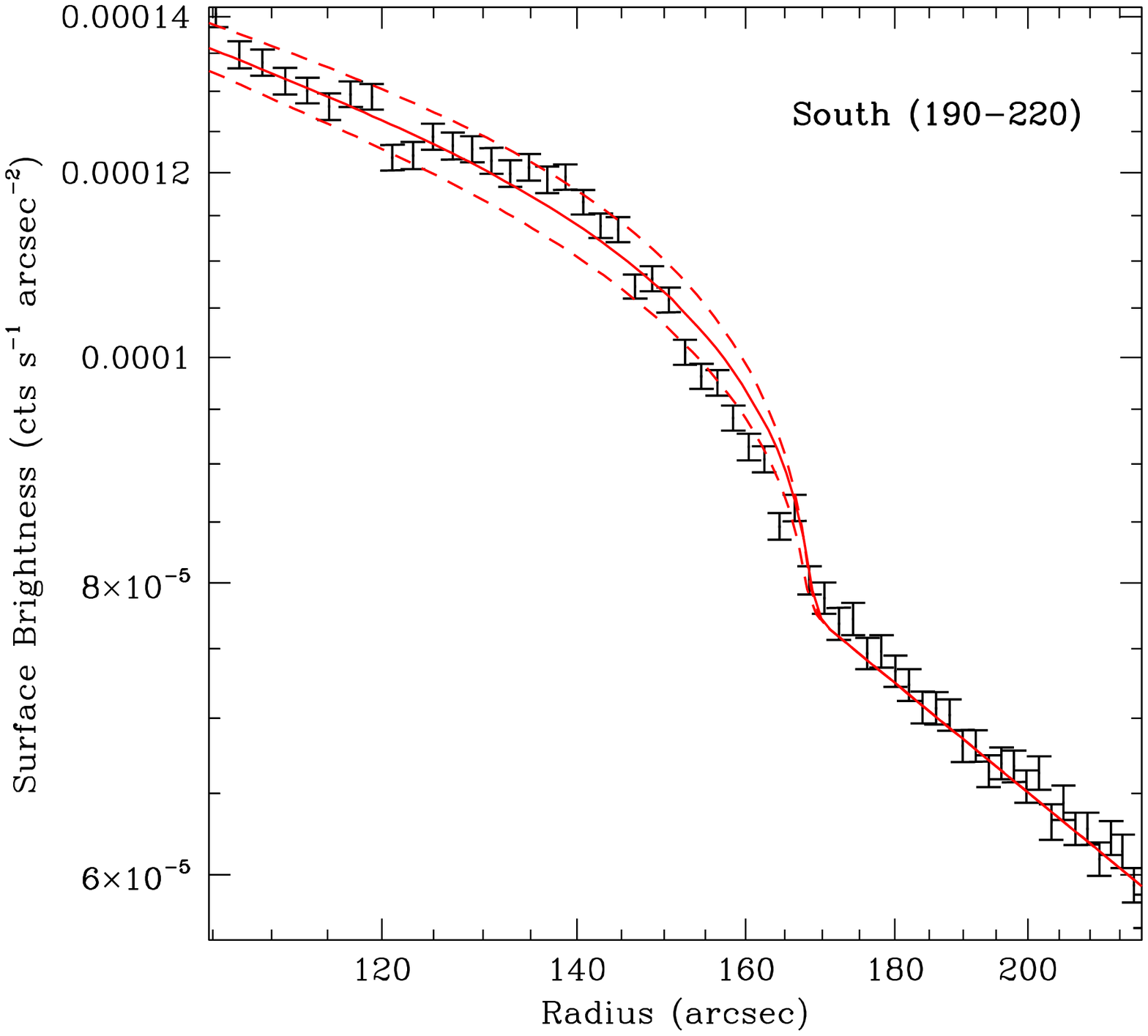}}\\
\scalebox{0.47}{\includegraphics[angle=270]{fig15c.ps}}
\hspace{0.4cm}
\scalebox{0.47}{\includegraphics[angle=270]{fig15d.ps}}
\caption{Projected surface brightness (top; in cts s$^{-1}$ arcsec$^{-2}$)
for position angles 190--220 (left) and position angles 350--20
(right).
The projected temperature profiles (bottom; in keV) 
are drawn from Fig. \ref{fig:kTFeNS} (left is south; right is north).
The surface brightness profiles show a clear discontinuity
at $\sim170-180$ arcsec, corresponding to the outer edge of the shock.
For the southern sector (position angles 190--220; left)
shock models with Mach numbers 1.25, 1.29, and 1.33 at a radius of 
168 arcsec are plotted; for the northern sector (position angles 350--20; 
right), 
models with Mach numbers
1.30, 1.34, and 1.38 at a radius of 180 arcsec are shown.
The overplotted spread in Mach numbers represent a conservative estimate for
the uncertainty in the measured Mach number.
For the temperature profiles, we approximate the ambient temperature profile
as linear fits to the temperature
profiles in the plotted radial range, excluding temperature bins
within the $\sim40$ arcsec thick shock.  The projected
temperature in the region of the shock is $\sim$5 per cent
hotter than the best fit ambient model.
}
\label{fig:vert}
\end{figure*}

Fig. \ref{fig:vert} shows the projected surface brightness
(upper panels; in cts s$^{-1}$ arcsec$^{-2}$)
for the position angles 190--220
(south of the AGN; left) and position angles 350--20 (north of the AGN; right).
These surface brightness profiles show signatures
commensurate with a weak shock at a radius of $r\sim180$ arcsec.
The relevant features occur at a slightly smaller 
radius to the south with respect to the north
($r\sim170$ as opposed to $r\sim180$ arcsec).
Overplotted on these profiles are simple 
shock models for a range of Mach numbers:
for the northern sector, Mach numbers of 1.30, 1.34 and 1.38; for 
the southern sector, Mach numbers 1.24, 1.29 and 1.33.

The shock models assume a spherically symmetric, hydrodynamic model of a point
explosion in an initially isothermal, hydrostatic atmosphere. 
The initial gas density profile is assumed to be a power law, which
is adjusted to match the observed surface brightness profile beyond the shock
(Nulsen \etal 2005a; Simionescu \etal 2009).
We have also explored a model that injects a constant amount
of energy per unit time 
into the system. This model provides no significant change to the
calculated surface brightness profile, which is already fit well by the
simple point explosion model.

These models show good overall agreement with the data and 
the analysis performed by Forman \etal (2007).
Table 2 shows the Mach number and shock radius 
from 4 different 30 degree wedges
(from -40--50 degrees and also from 190--220 degrees).
Systematic uncertainties dominate the modelling of the surface brightness
in these wedges. These include any projected non-sphericity of the shock
due to the likely jet axis alignment near our line of sight, and 
the incorrect assumption of an initially isothermal, hydrostatic atmosphere.
We, therefore, do not include standard,
statistical error bars from the models. Instead, we estimate the uncertainty
in the Mach number by finding a Mach number which brackets the observed
scatter in the surface brightness profiles
(see Fig. \ref{fig:vert}). This represents a conservative
estimate of the uncertainty in the Mach number of the shock.
The overall shape of the surface brightness constrains the Mach number of the
shock to within $\sim5$ per cent. 
The radius and Mach number of
the shock vary at the ten per cent level as a function of position angle.
This is likely due to the detailed structure of the ambient gas and shock
front in each sector.

\begin{figure}
\hspace{-0.3cm}
\scalebox{0.45}{\includegraphics{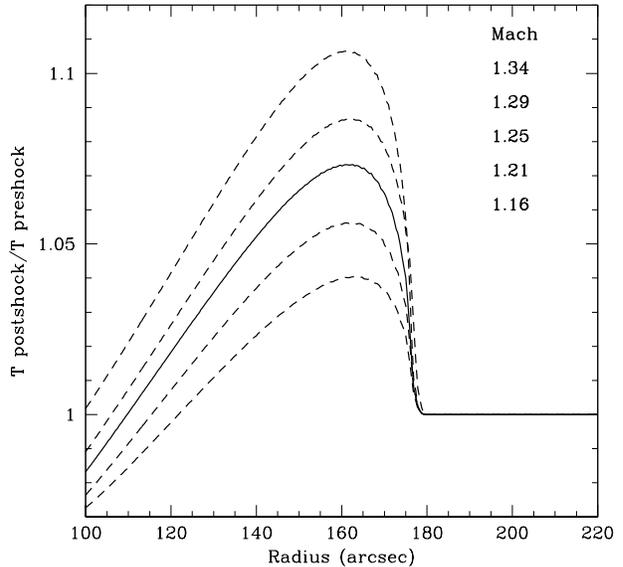}}
\caption{Expected emission weighted projected temperature profile divided
by the pre-shock temperature given the Mach numbers determined from
the shock models in 4 angular sectors. The average 
shock with Mach number $M=1.25$ at a radius of $r=180$ arcsec is represented
by the solid line.
The observed $\sim$5 per cent increase in temperature 
and the $\sim40$ arcsec thickness of the 
over-pressurized region is well explained by the shock model.
}
\label{fig:shock_kT}
\end{figure}

The temperature profiles for the position angles of $-135-86$ and $86-225$, 
drawn from Fig. \ref{fig:kTFeNS}, 
are shown in the lower panels of Fig. \ref{fig:vert} 
We have approximated the ambient temperature profile as a linear fit to the
plotted data range, excluding the bins within the $\sim40$ arcsec thick shock.
Under this assumption, we observe that both 
the northern and southern 
projected temperature profiles contain at least four bins 
(between $130\approxlt r\approxlt 170$
arcsec) that appear hotter ($\sim5$ per cent)
than expected.
Fig. \ref{fig:shock_kT} shows the expected projected temperature profile
divided by the pre-shock temperature profile as a function of radius. 
The $\sim$5 per cent rise in the temperature and the 
$\sim$40 arcsec thick shock in the 
observed, projected temperature profile are fully consistent with 
the shock model given an average Mach number $M=1.25$
(see Fig. \ref{fig:shock_kT}).
We, therefore, conclude that 
the high pressure ring is consistent with a shock with $M=1.25$.

Thick shocks have also been observed in the Perseus Cluster (see Fabian
\etal 2006), which does not show a clear 
temperature jump at the leading edge of the front.
The small increase in temperature within the thick 
over-pressurized ring in M\,87 suggests that 
shock heating is present in these systems but that 
the thickness and the small amount of heating associated 
with these weak shocks makes temperature increases 
difficult to detect.

\subsubsection{The inner shock}

A second shock is seen closer to the core ($\sim40$ arcsec; Section 3.2;  
Fig. \ref{fig:zoom})
The properties of this shock are difficult to constrain due to its
proximity to the core. Many other surface brightness features (\eg the
cavities filled with radio plasma seen in Fig.
\ref{fig:zoom}) are observed 
at similar radii, preventing an accurate determination
of the pre-shock gas properties and the
Mach number.  However, the projected 
temperature jump seen to the south of the AGN
at $r\sim40$ arcsec (3 kpc; Fig. \ref{fig:zoom}), 
suggests a Mach number of at least $M\approxgt1.2$.

The presence of the two shocks suggests that AGN outbursts are fairly
common. The estimated Mach numbers from these shocks and their
relative distances can be used to determine the frequency of these outbursts.
These shocks suggest that AGN outbursts occur approximately 
every $\sim10$ Myrs. Shock heating is likely only relevant in the central
regions of the cluster because shocks significantly weaken as they 
expand. If 
repeated shocks occur every $\sim10$ Myrs, as is suggested here, then
AGN driven, weak shocks could produce enough energy to offset the
radiative cooling of the ICM (Nulsen \etal 2007).

\subsection{Uplift of cool, metal-rich gas}
\label{section:pancake}

B\"ohringer \etal (1995; see also Churazov \etal 2001) 
propose that rising bubbles filled with radio
emitting plasma may be responsible for dragging cool, metal-rich
gas up out of the central regions of clusters.  
Buoyant bubbles will rise to a height that depends on their entropy,
stopping when they reach their appropriate adiabat.
This process may explain the observed radio edges at $r\sim300-400$ arcsec 
(23--31 kpc) and 
also at $r\sim400-500$ arcsec (31--39 kpc).  
When the bubbles reach their maximum height, 
they flatten.  A consequence
of this interpretation is that metal-rich gas rising with
the radio bubbles should be deposited at a radius similar to that at
which the
bubbles flatten.  Our abundance profiles show evidence
for enhancements in Fe abundance at $320<r<400$ arcsec (see Figs.
\ref{fig:kTFe}b, \ref{fig:kTFeNS}b). This is a similar radius to
the outer edges of the X-ray arms.  Indeed, bumps are observed
in the abundance profiles of all elements except Oxygen (Paper III).

The amount of Fe that is uplifted can be determined from the excess
metallicity measured at that radius and the gas mass.
In detail,

\begin{equation}
{M_{{\rm Fe}}=M_{{\rm g}}\times \Delta Z_{{\rm Fe}}\times f_{{\rm Fe}}},
\end{equation}

\noindent
where $M_{{\rm Fe}}$ and ${\Delta Z_{{\rm Fe}}}$ are the mass and 
excess abundance of
Fe, $M_{{\rm g}}$ is the gas mass, and $f_{{\rm Fe}}=\left<m_{{\rm Fe}}\right>/\left<m_{{\rm i\odot}}\right> \times \left( N_{{\rm Fe}}/\Sigma_{{\rm i}} N_{{\rm i}}\right)_{{\rm lodd}}$ is the mass
fraction of Fe (Lodders 2003). We measure the gas mass and excess
Fe abundance in the northern sector from a deprojection analysis
(using data from position angles of -108--55 degrees). This implies 
a total mass of uplifted Fe of $\sim1.0\times
10^6$ M$_{\odot}$. We compare this to the Fe mass currently present
in the arms (Simionescu \etal 2008).  For a gas mass of 
$5\times10^8$ M$_{\odot}$ and an Fe abundance of 2.2, those authors measure an 
Fe mass of $1.5\times10^6$ M$_{\odot}$ in the X-ray bright, cool arms.
Thus, a single generation of uplifted metals from buoyant bubbles is,
in principle, enough to
explain the observed excess Fe mass at large radius.  
It is unknown how much of the 
uplifted metals will remain at large radius after the bubbles flatten.
Lower significance bumps at higher radius 
that coincide with previous 
generations of radio bubbles may argue that this process occurs over multiple
generations.
It is likely that
the magnetic field configuration and its evolution will also have an impact
(\eg Bogdanovi\'c \etal 2009).
In the Perseus Cluster, 
metallicity enhancements also correspond well to the edges of 
several radio lobes, which suggests a similar origin
to that discussed here (see Sanders \etal 2005).
This process may also partly explain the metallicity ridges 
observed around the central regions of some cool core galaxy clusters
(Sanders \etal 2005; Sanders \etal 2009; Million \etal 2010).

We note that bumps in the Fe abundance profiles are not clearly
observed to the south. However, the mean 
Fe abundance is larger in the south than in 
the north and a cold front is located
at a similar radius (Simionescu \etal 2007, 2010), making the identification of
such a bump more difficult.
Our results
support the rising and pancaking bubble simulations presented in Churazov
\etal (2001; see also \eg Br\"uggen \& Kaiser 2002; Kaiser 2003; Br\"uggen
2003; Ruzkowski \etal 2004a,b; Heinz \& Churazov 2005).

\section{Summary}

Utilizing a detailed, spatially-resolved spectral mapping and
an ultra-deep (574 ks) {\it Chandra} observation of M\,87 and the
central regions of the Virgo Cluster, we present
an unprecedented, close-up view of AGN feedback in action.
Our maps reveal X-ray bright arms that have been lifted up by buoyant
radio bubbles as relatively cool, low entropy features
that are rich in structure (see \eg Belsole \etal 2001; Molendi 2002).
The detailed properties of these arms are discussed in Paper I.
Outside the arms, the ambient cluster gas is strikingly
isothermal at each radius. This suggests that either the gas remains relatively
undisturbed by AGN uplift or that conduction is efficient
along the azimuthal direction, as expected under action of the heat-flux
buoyancy instability (HBI).
Many cavities are seen in the inner radio cocoon ($\sim40$ arcsec or $\sim3$
kpc) including one located at $\sim7$ arcsec ($\sim0.5$ kpc) from the
central AGN in the counter-jet direction. These cavities are associated
with decreased surface brightness, high temperatures, low thermal
pressure, and typically appear filled with radio plasma in projection.

Our pressure map confirms spectrally previous results  
on the presence of a circular shock front at $r\sim180$ arcsec (14 kpc).
Based on the surface brightness jump, 
we measure an average Mach number of $M=1.25$ for the shock at 
$r\sim180$ arcsec, in good agreement
with the analysis of Forman \etal (2007).
Under the assumption of a linearly rising temperature profile,
we also measure a temperature jump within the $\sim40$ arcsec thick shock,
consistent with the expectations for a shock with Mach number 
$M=1.25$.
The thickness and overall weakness of these shocks will make it difficult
to detect temperature jumps unambiguously in hotter clusters like Perseus 
(\eg Fabian \etal 2006; Graham \etal 2008).

Another shock, associated with the central radio cocoon, is observed
at $\sim40$ arcsec ($\sim3$ kpc). Because of the complexity of this region,
we can only estimate the Mach number of this shock, using
the observed projected temperature jump across the front. This suggests
a Mach number of $M\approxgt1.2$.
The relative distances and Mach numbers of these shocks suggest
that AGN outbursts are common and occur every $\sim10$ Myrs.

Our results show that a significant mass of metals is lifted up to large
radius in the wake of the radio bubbles.
At least 
$1.0\times10^6 {\rm M_\odot}$ of Fe is uplifted to a radius 
of $350-400$ arcsec (27--31 kpc) to the north of the AGN.  
Given that a similar amount of Fe ($1.5\times10^6 {\rm M_\odot}$) is present
in the X-ray bright arms (Simionescu \etal 2008), 
a single generation of bubbles can, in principle, explain the observed 
Fe excess.
It is still uncertain how much metal
falls towards the central regions after this process occurs.  
Fe uplift of this magnitude may explain the observed 
metallicity ridges
observed around some cool core clusters. Studies of such
abundance features can further enhance our understanding of
transport processes. This will be the subject of future work.

\section{Acknowledgments}

We thank W.R. Forman, C. Jones, and E. Churazov for helpful comments
and R.G. Morris for computational support.
N. Werner and A. Simionescu 
were supported by the National Aeronautics and Space Administration 
through Chandra/Einstein Postdoctoral Fellowship Award Number PF8-90056 and 
PF9-00070 issued by the Chandra X-ray Observatory Center, which is operated 
by the Smithsonian Astrophysical Observatory for and on behalf of the National 
Aeronautics and Space Administration under contract NAS8-03060. This work was 
supported in part by the US Department of Energy under contract number 
DE-AC02-76SF00515. All computational analysis was carried out using the
KIPAC XOC compute cluster at Stanford University and the Stanford
Linear Accelerator Center (SLAC).

\end{document}